\documentclass[aps,prd,preprint,superscriptaddress,showpacs,floatfix,nobibnotes,nofootinbib]{revtex4-1}

\usepackage{float}

\usepackage{latexsym}
\usepackage{amsmath}
\usepackage{amssymb}
\usepackage{graphicx}
\usepackage{setspace}
\usepackage{longtable}
\usepackage{color}
\usepackage{enumitem}
\usepackage[bottom]{footmisc}

\usepackage{bm}
\usepackage{epsfig}
\newcommand{\bea}{\begin{eqnarray}}
\newcommand{\eea}{\end{eqnarray}}

\newcommand{\nn}{\nonumber}
\newcommand{\ww}{{\rm w}}

\begin{document}
\setlength{\baselineskip}{19pt}

\title{The energy-momentum tensor at the earliest stage of relativistic heavy-ion collisions}

\author{Margaret E. Carrington}
\affiliation{Department of Physics, Brandon University,
Brandon, Manitoba R7A 6A9, Canada}
\affiliation{Winnipeg Institute for Theoretical Physics, Winnipeg, Manitoba, Canada}

\author{Alina Czajka}
\affiliation{National Centre for Nuclear Research, ul. Pasteura 7,  PL-02-093  Warsaw, Poland}

\author{Stanis\l aw Mr\' owczy\' nski}
\affiliation{Institute of Physics, Jan Kochanowski University, ul. Uniwersytecka 7, PL-25-406 Kielce, Poland }
\affiliation{National Centre for Nuclear Research, ul. Pasteura 7,  PL-02-093  Warsaw, Poland}

\date{September 13, 2021}

\begin{abstract}
Nuclear collisions at high energies produce a gluon field that can be described using the Colour Glass Condensate (CGC) effective theory at proper times $\tau \lesssim 1$ fm. 
The theory can be used to calculate the gluon energy-momentum tensor, which provides information about the early time evolution of the chromo-electric and chromo-magnetic fields, energy density, longitudinal and transverse pressures, and other quantities. 
We obtain an analytic expression for the energy-momentum tensor using an expansion in the proper time, and working to sixth order. 
The calculation is technically difficult, in part because the number of terms involved grows rapidly with the order of the $\tau$ expansion, but also because of several subtle issues related to the definition of event-averaged correlators, the method chosen to regulate these correlators, and the dependence of results on the parameters introduced by the regularization and nuclear density profile functions. 
All of these issues are crucially related to the important question of the extent to which we expect a CGC approach to be able to accurately describe the early stages of a heavy-ion collision. 

We present some results for the evolution of the energy density and the longitudinal and transverse pressures.
We show that our calculation gives physically meaningful results up to values of the proper time which are close to the regime at which hydrodynamic simulations are initialized. 
In a companion paper \cite{carrington1} we give a detailed analysis of several other experimentally relevant quantities that can be calculated from the energy-momentum tensor.

\end{abstract}

\maketitle

\section{Introduction}

The study of quark-gluon plasma (QGP) is of fundamental interest because it is a means to obtain information about the structure of the phase diagram of quantum chromodynamics (QCD). 
The temporal evolution of the matter produced in relativistic heavy-ion collisions is reasonably well understood in general terms, but there is not a lot of information about the details of its evolution during the earliest stage. 
There are no directly accessible experimental signals of this phase, and most information about its features is lost by the time equilibrium is reached. 
Exceptions are electroweak and hard probes which are sensitive to the full history of the spacetime evolution of the system, but are very difficult to measure or interpret. 
The  information about the system's non-equilibrium evolution that survives, is imprinted as initial conditions for its subsequent hydrodynamic evolution. 
The development of a quantitative description of the dynamics of the highly non-equilibriated early-time system would allow us to calculate the early-time distributions of energy and momentum, which would provide the initial conditions that are the required input parameters of hydrodynamical simulations. 

A great deal of effort has been invested in this difficult and important task. 

The Glauber model was one of the first attempts, see the review \cite{Miller:2007ri}. 
The model is based on a quantum-mechanical description of the collision in terms of geometric parameters such as 
the impact parameter and the distribution of nucleons in the colliding nuclei, as well as some phenomenological constraints. 
Calculations of observable quantities require sophisticated numerical techniques to perform high-cost multi-dimensional integrals. 
An alternative approach was developed in Refs. \cite{Kharzeev:2000ph,Kharzeev:2001gp} and is called the KLN model. 
The model uses the framework of high density QCD in which the system is described in terms of
parton saturation or, in the language of colour fields, using classical chromodynamics.
A scaling function is derived that describes the dependence of hadron multiplicities on energy, centrality, rapidity, and atomic
number. 
A Monte-Carlo implementation of the approach which includes high-rapidity fluctuations of hard sources was developed in Refs.~\cite{Drescher:2006ca,Drescher:2007ax}. 

The Colour Glass Condensate (CGC) is an effective theory that describes high energy nucleons and nuclei \cite{McLerran:1993ni, McLerran:1993ka, McLerran:1994vd}. 
The theory is based on a separation of scales between source partons with large nucleon momentum fraction (denoted x), and classical gluon fields with small nucleon momentum fraction. 
When the separation scale is fixed, the dynamics of the small x gluons can be determined from the classical Yang-Mills (YM) equation with the source provided by the large x valence partons.
One then averages over an ensemble of source colour charge densities. 
The original McLerran-Venugopalan (MV) model assumed a source charge density that was infinitely Lorentz contracted and homogeneous in the transverse plane, but it was later realized that it is necessary to include a finite width of the sources across the light-cone \cite{JalilianMarian:1996xn, Kovchegov:1999yj}.

The gluonic state that is produced at very early times after a relativistic heavy-ion collision is called a glasma. 
The classical gluon fields of the individual nuclei before the collision, and the glasma fields in the post-collision system, satisfy the YM equation. Boundary conditions connect 
the known pre-collision solutions to the glasma field immediately after the collision. 
In the post-collision regime, one can consider a systematic expansion of the YM equation using a power series in the proper time $\tau$. 
The idea of performing an analytic $\tau$ expansion, also called the ``near field expansion,'' was proposed in \cite{Fries:2005yc} and further developed in \cite{Fukushima:2007yk,Fujii:2008km,Chen:2015wia,Fries:2017ina}. 
The convergence of the series is determined by the time scale  $\tau_0\sim 1/Q_s$, where $Q_s$ is the saturation scale. 
We note again that although $\tau_0$ is a very early time, it may be sufficient to determine important bulk properties of the plasma, including the distributions of the energy, momentum, and angular momentum that are transferred from the initial colliding nuclei to the glasma around mid-rapidity. 

A modification of this approach called the IP-Glasma model was studied in Refs.~\cite{Schenke:2012wb,Schenke:2012hg}.
The IP-Glasma model combines the Impact Parameter Saturation (IP-Sat) model \cite{Kowalski:2003hm,Watt:2007nr}, which is a generalization of the MV model that describes finite systems by promoting the colour
charge density to depend on the impact parameter,  and the evolution of glasma fields through classical YM equations.  
The essential input is the dipole cross section in proton deep inelastic scattering, which is experimentally well constrained. 
The result of the procedure is a ``lumpy'' distribution for the saturation scale $Q_s(x^\pm,\vec x_\perp)$  which describes the sub-nucleon structure of the nucleus. Using this result, one obtains a  good description of many bulk features of distributions at RHIC and the LHC \cite{Schenke:2013dpa,Schenke:2014tga}.
The original formulation assumed boost invariance and was therefore  effectively 1+2 dimensional, but fluctuations in rapidity have since been included using the JIMWLK renormalization group equation, see~\cite{Schenke:2016ksl,McDonald:2018wql,McDonald:2020oyf}, and provide a full 1+3 dimensional glasma picture. 
Boost non-invariant systems have also been studied using coloured ``particle-in-cell" simulations \cite{Gelfand:2016yho,Ipp:2017lho,Ipp:2020igo}. 
These calculations simulate the evolution of coloured point charges in continuous phase space coupled to non-Abelian gauge fields on a discrete lattice.

We perform a fully analytic calculation of the energy-momentum tensor using a near field expansion, working to sixth order in $\tau$, and including  full dependence on rapidity and the source charge densities of each ion. 
The resulting expression is so long that we will not explicitly write all terms in this paper. 
We stress however that an advantage of obtaining an analytic result, no matter how lengthy, is that there are no issues with possible errors introduced by a discretized numerical procedure. 
There are several subtle issues associated with the calculation of correlation functions and the regularization of these functions, and  different approaches appear in the literature. We discuss these issues in some detail, and make some comparisons with methods employed by different authors. 
We also discuss the validity of the near field expansion. 
The idea of expanding in the proper time was proposed almost 15 years ago but there are only a few calculations in the literature that make use of the method, 
and its convergence has never been studied. We compare results at different orders in the expansion and show that the sixth order expressions are reliable to about $\tau \sim 0.05$ fm. 
We also address the question of whether or not the classical picture that is inherent in the formulation of the CGC approach we are using, is valid for these short times. 
We show that the region of validity of the near field expansion reaches far beyond the lower bound at which we no longer trust the classical description we are using.
Finally, we comment that analytic solutions of the near field expanded YM equation are useful in other contexts. 
As one example, we have developed a method to calculate transport properties of heavy quarks in glasma \cite{Mrowczynski:2017kso,Carrington:2020sww}, and we are currently extending these calculations to higher orders in the $\tau$ expansion using the results presented in this paper.

We comment that in spite of its phenomenological success, the CGC approach that we use does not capture completely the QCD dynamics
of a heavy ion collision.  The method has been developed in various directions, see for example references \cite{Balitsky:2004rr, Hatta:2005rn,Chirilli:2015fza} and the review \cite{Gelis:2012ri}.

This paper is organized as follows. 
In section \ref{sec-preliminaries} we describe some parts of the CGC effective theory that are relevant to our work, and define our notation.
In section \ref{sec-method} we give some details of our calculation of the energy-momentum tensor. 
At sixth order in $\tau$ the energy-momentum tensor contains a very large number of terms, and the calculation was therefore done with the help of {\it Mathematica}. 
The complete result is too long to give explicitly, but is available on request to interested readers in the form of a {\it Mathematica} or text file.
We describe some of the many checks we have performed to verify the result. 
In section \ref{sec-reg} we discuss our method to regulate the correlators of gauge potentials which enter our expression for the energy-momentum tensor.
In section  \ref{sec-results} we give some parts of our analytic expression for the energy-momentum tensor and present some numerical results.
In this paper we show results only for the special case of nuclei with infinite extent in the transverse plane that are invariant under translations and rotations in the plane. 
In our companion paper \cite{carrington1} we present a detailed analysis of various experimentally relevant quantities that can be calculated from our full results for the energy-momentum tensor at order $\tau^6$. In section \ref{sec-conc} we make some concluding remarks.

\section{The CGC effective theory}
\label{sec-preliminaries}
In this section we discuss the formalism of the CGC effective theory that we use and present our notation. Further notational details and a collection of some useful formulas can be found in Appendix \ref{AppendixA}. 

We consider a collision of two heavy ions moving towards each other along the $z$-axis and colliding at $t=z=0$. 
The transverse degrees of freedom are denoted by the 2-vector $\vec x_\perp$. 
The time and longitudinal coordinate ($t,z)$ can be written in two different combinations which will both be  useful: in different situations we use either light-cone coordinates, $x^\pm=(t\pm z)/\sqrt{2}$, or Milne coordinates, $\tau=\sqrt{t^2-z^2}=\sqrt{2x^+ x^-}$ and $\eta=\ln(x^+/x^-)/2$.

Tensor equations, like equation (\ref{YMeqn}), are valid in any coordinate system. 
However, in some parts of our calculation it will be easier to use a particular basis. 
All vectors and tensors can be written in the Minkowski, light-cone or Milne basis. 
For example: we can write either $A_{\rm mink}^\mu(x)\equiv(A^0(x),A^z(x),\vec A(x))$, $A_{\rm lc}^\mu(x)\equiv(A^+(x),A^-(x),\vec A(x))$ or $A_{\rm milne}^\mu(x)\equiv(A^\tau(x),A^\eta(x),\vec A(x))$. 
Transformations from one basis to another are performed using the appropriate general coordinate transformation (see equation (\ref{milne2mink})). 
The  transverse components of any vector or tensor are the same in all three bases, and we will use indices $(i,j,k,l\dots)$ to denote  transverse components. Individual components are sometimes written with letter indices, using an obvious notation (for example, $A^{i=1} \equiv A^x$ and $A^{i=2} \equiv A^y$).
In most cases we use a specific choice of basis consistently within a given section of this paper.
In addition, in most equations it is obvious which basis is being used (for example in equation (\ref{ansatz}) the superscripts on the left side make it clear that the potential is written in the light-cone basis). 
In any situation where the basis is not clear, we include a subscript stating explicitly which basis is used. 

In the formulation of the CGC effective theory that we use, the dynamics of the small x gluons is determined from the classical YM equation
\bea
\label{YMeqn}
[D_\mu,F^{\mu\nu}]=J^\nu\,
\eea
where 
\bea
F_{\mu\nu}=\frac{i}{g}[D_\mu,D_\nu]\,\text{~~and~~} \, D_\mu = \partial_\mu -igA_\mu\,.\label{field-strength}
\eea
Both $J^\mu$ and $A^\mu$ are SU($N_c$) valued functions that can be written as linear combinations of the group generators $t_a$, for example $A^\mu=A^\mu_a t_a$. The generators satisfy $[t_a,t_b]=if_{abc}t_c$ where $f_{abc}$ are the structure constants of SU$(N_c)$.
The two ions moving towards each other along the $z$-axis contain large x valence partons that provide the source on the right side of equation (\ref{YMeqn})
\bea
\label{J-LC}
&& J^\mu(x) = J^{\mu}_1(x)+J^{\mu}_2(x) \\
&& J^{\mu}_1(x)=\delta^{\mu +} g \rho_1(x^-,\vec{x}_\perp)\text{~~and~~}J^{\mu}_2(x)=\delta^{\mu - } g \rho_2(x^+,\vec{x}_\perp)\,\nonumber
\eea
where the indices 1 and 2 indicate the ions moving to the right (the positive $z$ direction) and left (the negative $z$ direction), respectively. 
The partons are assumed to remain ultra-relativistic throughout the collision, and the 
currents are static (independent of the light-cone time). Physically this means that the lifetime of the valence partons is  much greater than that of the small x degrees of freedom. We refer to the path of ion 1 as the positive light-cone, and ion 2 moves along the negative light-cone. 
Because of Lorentz contraction, both ions have a very small but finite region of support across the light-cone over $-\ww/2\le x^\mp \le \ww/2$. 
The limit ${\rm w}\to 0$ will be taken at the end of the calculation (see Appendix \ref{BBxx} for details), but in intermediate steps of the calculation it is necessary to keep w non-zero. 

Our goal is to find the energy-momentum tensor in the forward light-cone, which corresponds to the post-collision part of spacetime. It is natural to describe this region using Milne coordinates. We will work in the gauge $A^\tau=0$, which is called axial gauge, or Fock-Schwinger gauge. 
The axial gauge potential in the forward light-cone has the simple form
\bea
\label{ansatz2}
A^\mu_{\rm milne} =\theta(\tau) \big(0,\alpha(\tau, \vec x_\perp),\vec\alpha_\perp(\tau, \vec x_\perp)\big)\,
\eea
where the functions $\alpha(\tau,\vec x_\perp)$ and $\vec\alpha_\perp(\tau,\vec x_\perp)$ are independent of rapidity, as is appropriate for a boost invariant system. The YM equation and the energy-momentum tensor take a simple form in these coordinates. 
As will be explained in detail below, our method is to expand in $\tau$, solve the YM equation order by order in the $\tau$ expansion, and obtain expressions that depend on the initial potentials $\alpha(0, \vec x_\perp)$ and $\vec\alpha_\perp(0, \vec x_\perp)$. 

The initial potentials must then be connected to the source terms that represent the currents of the two colliding ions. 
To do this, we need to write the gauge potential in light-cone coordinates, where we can separate the regions of spacetime that correspond to the pre- and post-collision fields. 
Using the general coordinate transformation in (\ref{milne2mink}), one finds that in light-cone coordinates the axial gauge condition $A^\tau=0$  takes the form  
\bea
\label{foch}
x^+ A^- + x^- A^+  = 0\,.
\eea
We use the following ansatz for the gluon potential \cite{Kovner:1995ts, Kovner:1995ja}
\bea
\label{ansatz}
&& A^+(x) = \Theta(x^+)\Theta(x^-) x^+ \alpha(\tau,\vec x_\perp)  \\
&& A^-(x) = -\Theta(x^+)\Theta(x^-) x^- \alpha(\tau,\vec x_\perp) \nonumber\\
&& A^i(x) = \Theta(x^+)\Theta(x^-) \alpha_\perp^i(\tau,\vec x_\perp)
+\Theta(-x^+)\Theta(x^-) \beta_1^i(x^-,\vec x_\perp)
+\Theta(x^+)\Theta(-x^-) \beta_2^i(x^+,\vec x_\perp)\,\nonumber
\eea
which satisfies equation (\ref{foch}) in all regions of spacetime.
The theta functions separate the glasma potentials in the post-collision part of  spacetime, determined by the  functions $\alpha$ and $\vec\alpha_\perp$, from the pre-collision potential of each ion, denoted $\vec\beta_1$ and $\vec\beta_2$.
In the post-collision part of spacetime, the four-component vector potential is represented in terms of three independent scalar functions.
In each of the pre-collision regions there are only two independent functions, which can be thought of as a convenient use of residual gauge freedom.

\section{Method} 
\label{sec-method}
Our ultimate goal is to calculate the energy-momentum tensor in the post-collision region of spacetime. 
There are four main steps:
\begin{enumerate}[label=\Alph*.]

\item 
The post-collision gauge potentials satisfy the YM equation (\ref{YMeqn}) with the source term on the right side set to zero. 
We work in Milne coordinates with the ansatz (\ref{ansatz}),  expand the functions $\alpha(\tau,\vec x_\perp)$ and $\vec\alpha_\perp(\tau,\vec x_\perp)$ to sixth order in $\tau$, and solve the YM equation for the coefficients of these expansions. 

\item Using the solutions obtained in step A, we find the energy-momentum tensor in the post-collision region in terms of the initial potentials 
$\alpha(0,\vec x_\perp)$ and $\vec\alpha_\perp (0,\vec x_\perp)$ and their derivatives. We perform a coordinate transformation to convert  our result for the energy-momentum tensor from the Milne basis to the Minkowski basis. 

\item We  apply the boundary conditions that connect the  potentials $\alpha(0,\vec x_\perp)$ and $\vec\alpha_\perp(0,\vec x_\perp)$ to the potentials $\vec\beta_1(x^-,\vec x_\perp)$ and $\vec\beta_2(x^+,\vec x_\perp)$ in the pre-collision region. Using these results we rewrite the Minkowski space energy-momentum  tensor that was obtained in step B in terms of pre-collision potentials for each ion, and their derivatives. 

\item The pre-collision potentials $\beta^i_1(x^-,\vec x_\perp)$ and $\beta^i_2(x^+,\vec x_\perp)$ are generated by the individual incoming nuclei and we find these potentials by solving the YM equation in the pre-collision region. 
These solutions depend on the source charge distributions (\ref{J-LC}), which are not known.
An important input to the CGC approach is the assumption of a Gaussian distribution of colour charges within each nucleus. 
One then calculates correlation functions of pre-collision potentials by averaging over these Gaussian distributions.
We rewrite the energy-momentum tensor obtained in step C in terms of correlations of pre-collision potentials, and calculate these correlators by averaging over the source distributions.  
\end{enumerate}

\subsection{Forward light-cone potentials}
\label{sec-forward-solutions}

{\it All equations in this section are written in Milne coordinates.}

In this section we solve the sourceless YM equation in the forward light-cone in Milne coordinates. 
The sourceless YM equation (\ref{YMeqn}) can be written
\bea
\label{YM-0}
g^{\sigma \mu}\big[\nabla_\sigma[\nabla_\mu,\nabla_\nu]\big]=0\,
\eea
where the covariant derivative in this equation includes both the gauge field contribution, and the Christoffel symbols that describe the curvature in Milne coordinates (see equations (\ref{nabla}, \ref{connect})). 
The ansatz for the gauge potential (\ref{ansatz}) 
contains only three functions $\alpha$ and $\vec\alpha_\perp$ in the forward light-cone (due to the gauge condition), and therefore the four components of the YM equation are not all independent. 
We use the three equations obtained from setting $\nu\in(1,2,3)$ 
and find analytic solutions for the three ansatz functions  by expanding in the proper time  and solving for the coefficients of the expansion. 
We write
\bea
\alpha(\tau,\vec x_\perp) &=& \alpha^{(0)}(\vec x_\perp) + \tau \alpha^{(1)}(\vec x_\perp) + \tau^2 \alpha^{(2)}(\vec x_\perp) + \cdots = \sum_{n=0} \tau^n \alpha^{(n)}(\vec x_\perp) \label{exp1}
\eea
and similarly
\bea
\vec\alpha_\perp(\tau,\vec x_\perp) &=& \sum_{n=0} \tau^n \vec\alpha^{(n)}_\perp(\vec x_\perp) \,.\label{exp2}
\eea
It can be shown using a recursion relation that all of the coefficients multiplying odd powers of $\tau$ are zero \cite{Chen:2015wia}.
The coefficients $\alpha^{(2)}(\vec x_\perp)$ and $\vec\alpha_\perp^{(2)}(\vec x_\perp)$ are found in terms of $\alpha^{(0)}(\vec x_\perp)$ and $\vec\alpha^{(0)}(\vec x_\perp)$, the coefficients $\alpha^{(4)}(\vec x_\perp)$ and $\vec\alpha_\perp^{(4)}(\vec x_\perp)$ are then found in terms of $\alpha^{(2)}(\vec x_\perp)$, $\vec\alpha_\perp^{(2)}(\vec x_\perp)$, $\alpha^{(0)}(\vec x_\perp)$ and $\vec\alpha_\perp^{(0)}(\vec x_\perp)$, etc. 
The process is tedious but perfectly straightforward and can be carried out to any order using {\it Mathematica}. 
As a consistency check, we verify that the solution obtained satisfies the YM equation with $\nu=0$.

The results can be written in compact form in terms of the fields at lowest order in the $\tau$ expansion. 
The only non-zero components of the electric and magnetic fields at $\tau=0$ are
\bea
&& E \equiv E^z(0,\vec x_\perp) = -2 \alpha(0,\vec x_\perp)\nn\\
&& B \equiv B^z(0,\vec x_\perp) = \partial^y \alpha_\perp^x(0,\vec x_\perp) - \partial^x \alpha_\perp^y(0,\vec x_\perp) -ig[\alpha_\perp^y(0,\vec x_\perp),\alpha_\perp^x(0,\vec x_\perp)]\,.\label{Bz-initial}
\eea
To fourth order, the even coefficients in the series in equations (\ref{exp1}, \ref{exp2}) are (omitting all arguments)
\begin{eqnarray}
\label{al-2}
&& \alpha^{(2)} = \frac{1}{8} [{\cal D}^j,[{\cal D}^j,\alpha^{(0)}]] \nn\\
\label{al-4}
&& \alpha^{(4)} = \frac{ig}{48} \epsilon^{ij} \left[[{\cal D}^i, \alpha^{(0)}],[{\cal D}^j, B]\right] 
+\frac{1}{192} [{\cal D}^k,[{\cal D}^k,[{\cal D}^j,[{\cal D}^j,\alpha^{(0)}]]]] \nn\\
\label{al-p2}
&& \alpha^{i(2)}_{\perp} = \frac{1}{4} \epsilon^{ij} [{\cal D}^j, B]\nn \\
\label{al-p4}
&& \alpha^{i(4)}_{\perp} = \frac{ig}{64}  \left[\left[{\cal D}^i,B\right],B\right] 
+ \frac{1}{64}\epsilon^{ij} \left[{\cal D}^j,\left[{\cal D}^k, \left[{\cal D}^k,B \right]\right]\right] 
 + \frac{ig}{16}\left[\alpha^{(0)},\left[{\cal D}^i_,\alpha^{(0)}\right]\right] \,
\end{eqnarray}
where we define
\bea
&& {\cal D}^i = \partial^i-ig \alpha_\perp^i(0,\vec x_\perp)\,
\eea
and the notation $\epsilon^{ij}$ represents a matrix with values $\epsilon^{11}=\epsilon^{22}=0$ and $\epsilon^{12}=-\epsilon^{21}=1$.

\subsection{Energy momentum tensor in the forward light-cone}
\label{sec-Tmunu-start}

The field-strength tensor is 
\bea
F_{\mu\nu} = \frac{i}{g}[\nabla_\mu,\nabla_\nu]\,.
\eea
Using equations (\ref{ansatz2}, \ref{exp1}, \ref{exp2}) and the solutions found in the previous section, we obtain a lengthy expression for the field-strength tensor in Milne coordinates that depends only on the initial potentials $\alpha^{(0)}(\vec x_\perp)$ and $\vec\alpha^{(0)}_\perp (\vec x_\perp)$ and the expansion parameter $\tau$. 

The energy-momentum tensor can be written in terms of the field-strength tensor  as
\bea
&& T^{\mu\nu} = 2{\rm Tr}\big[F^{\mu\lambda}F_\lambda^{~\nu}+\frac{1}{4}g^{\mu\nu}F^{\alpha\beta}F_{\alpha\beta}\big] 
= F_a^{\mu\lambda}F_{\lambda\,a}^{~\nu}+\frac{1}{4}g^{\mu\nu}F_a^{\alpha\beta}F_{\alpha\beta\,a}\,\label{Tmunu}
\eea
where we have used the definition obtained by adding a total divergence to the canonical result to produce an expression that is gauge invariant, conserved, symmetric and traceless. 
The two field-strength tensors in equation (\ref{Tmunu}) are actually calculated at different points, denoted $x=(x^+,x^-,\vec x_\perp)$ and $y=(y^+,y^-,\vec y_\perp)$, and we take $(x-y)\to 0$ at the end of the calculation. 
In this section we do not write these arguments explicitly. 
The entire calculation can be done with {\it Mathematica}.
We have checked our method by verifying that the resulting expression is symmetric and has zero divergence
\bea
\left[\partial_\mu T^{\mu\nu} - \Gamma^\mu_{\mu\alpha}T^{\alpha\nu}-\Gamma^\nu_{\mu\alpha}T^{\mu\alpha}\right]_{\rm milne} = 0\,.
\eea
The energy-momentum tensor in Minkowski space can then be found using the coordinate transformation $M$ given in equation (\ref{milne2mink}):
\bea
&& T^{\mu\nu}_{\rm mink} = M^\mu_{~\alpha} M^\nu_{~\beta} T^{\alpha\beta}_{\rm milne}\,.\label{TTmink}
\eea
A useful expression for the energy-momentum tensor in terms of the electric and magnetic fields is given in Appendix \ref{field-exp-BETA}.

\subsection{Boundary conditions}
\label{sec-initial-conditions}

The expression for the energy-momentum tensor obtained at the end of the previous section depends on the initial potentials $\alpha^{(0)}(\vec x_\perp)$ and $\vec\alpha^{(0)}_\perp (\vec x_\perp)$, which are related to the pre-collision potentials $\vec\beta_1(x^-,\vec x_\perp)$ and $\vec\beta_2(x^+,\vec x_\perp)$ through a set of boundary conditions. The energy-momentum tensor also depends on potentials that depend on the $y$ coordinates, because we have not yet taken the limit $(x-y)\to 0$, but everything in this section can be extended trivially to potentials that depend on $y$ instead of $x$.

We need to use boundary conditions that connect the initial potentials in the forward light-cone with those in the pre-collision regions.  
These conditions were originally obtained by matching terms from the pre- and post-collision regions that are singular on the light-cone \cite{Kovner:1995ts, Kovner:1995ja}. We work with sources with small but finite width, and the limit that this width goes to zero cannot be taken until after the boundary conditions have been used. These boundary conditions should therefore be obtained by integrating the YM equation across the light-cone. 
Some details of this calculation are given in Appendix \ref{initial-sec}. The boundary conditions are
\bea
\label{cond1}
&& \alpha^{i}_\perp(0,\vec{x}_\perp) = \alpha^{i(0)}_\perp(\vec{x}_\perp) = \lim_{\text{w}\to 0}\left(\beta^i_1 (x^-,\vec{x}_\perp) + \beta^i_2
(x^+,\vec{x}_\perp)\right)\nonumber \\
&& \label{cond2}
\alpha(0,\vec{x}_\perp) = \alpha^{(0)}(\vec{x}_\perp) = -\frac{ig}{2}\lim_{\text{w}\to 0}\;[\beta^i_1 (x^-,\vec{x}_\perp),\beta^i_2
(x^+,\vec{x}_\perp)]\,\label{b-conds}
\eea
where the notation $\lim_{\text{w}\to 0}$ indicates that the width of the sources across the light-cone is taken to zero. As explained in section \ref{sec-corr}, the pre-collision potentials depend only on transverse coordinates in this limit. Using equation (\ref{cond2}) it is straightforward to rewrite the energy-momentum tensor obtained at the end of section \ref{sec-Tmunu-start} in terms of pre-collision potentials only.

\subsection{Correlation functions of pre-collision potentials}
\label{sec-corr}

The result obtained at the end of the procedure outlined in section \ref{sec-initial-conditions} is a lengthy expression for the energy-momentum tensor in terms of products of pre-collision potentials. 
A generic term  has the form 
\bea
\beta_{1a}^i(x^-,\vec x_\perp) \beta_{1b}^k(x^-,\vec x_\perp)\beta_{2c}^k(x^+,\vec x_\perp) \dots \beta_{1d}^m(y^-,\vec y_\perp)\beta_{2e}^l(y^+,\vec y_\perp) \beta_{2f}^j(y^+,\vec y_\perp)  \dots  \,
\label{core-ex-1}
\eea
The pre-collision  potentials can be expressed in terms of the ion sources by solving the YM equation in the pre-collision region.
%We can write the pre-collision potentials in terms of the source functions $\rho_1(x^-, \vec x_\perp)$ and $\rho_2(x^+, \vec x_\perp)$ (some details are given in Appendix \ref{BBxx}). THH  
These colour charge distributions are not known, and an important input to the CGC approach is the use of an averaging procedure based on the assumption of a Gaussian distribution of colour charges within each nucleus. 
A product of colour charges is replaced by its average over this Gaussian distribution (which will be denoted with angle brackets).
We make the assumption that sources from different ions are un-correlated, or equivalently that correlation functions of products of sources from different ions can be set to zero. 
This means that we need to consider only averages of the form $\langle \rho_1 \rho_1 \dots \rho_1\rangle$ and $\langle \rho_2 \rho_2 \dots \rho_2\rangle $ where all sources are from the same ion. 
This approximation is justified because the pre-collision distributions are independent from each other due to causality, 
and they remain uncorrelated post-collision because the CGC approach we are using assumes that the color sources are static. 
We comment that the assumption of static sources is a standard component of most CGC calculations, but it necessarily means that the method does not capture the QCD dynamics of a heavy ion collision completely.

Local fluctuations are assumed proportional to the colour charge density. For the first ion we denote the colour charge density $\lambda_1(x^-,\vec x_\perp)$ and define 
\bea
\label{big-in}
\langle \rho_1
(x^-,\vec x_\perp)
\rho_1(y^-,\vec y_\perp)\rangle 
\equiv \,g^2\,\lambda_1(x^-,\vec x_\perp)\delta(x^--y^-)\delta^2(\vec x_\perp - \vec y_\perp)\,
\eea
where $\lambda_1(x^-,\vec x_\perp)  = h(x^-) \mu_1(\vec x_\perp) $ with $h(x^-)$ a sharply peaked non-negative function with width ${\rm w}$  around $x^-=0$ which is normalized to one. Integration gives
\bea
 \int dx^-\,\lambda_1(x^-,\vec x_\perp)=\int dx^-\,h(x^-) \,\mu_1(\vec x_\perp) \equiv \mu_1(\vec x_\perp)\,
 \label{norm}
\eea
where $\mu_1(\vec x_\perp)$ is a surface colour charge density. We make the analogous definitions for the second ion, and the width w is taken to zero at the end of the calculation. 
The average over a Gaussian distribution of colour densities which are independent random variables can be rewritten as a sum over the averages of all possible pairs, a result known as Wick's theorem, and therefore the average over any product of colour sources can be written in terms of the fundamental correlator in equation (\ref{big-in}).

A product of pre-collision potentials, like the generic term in equation (\ref{core-ex-1}), must be expressed as some average over a product of source functions.
Products of pre-collision potentials can be written as products of Wilson lines and source functions (see Appendix \ref{BBxx} for details). 
The Gaussian averaging of products of this form have been studied by many authors 
\cite{Blaizot:2004wv, Fukushima:2007dy, Albacete:2018bbv, FillionGourdeau:2008ij, Lappi:2017skr}. 
These calculations are very involved, and the difficulties increase rapidly as the number of potentials grows. 

In this paper we use the approximation that Wick's theorem can be applied to light-cone potentials directly, which is sometimes called the Glasma Graph approximation \cite{Lappi:2015vta}. 
Correlations of any even number of potentials from the same ion are written as products of correlators of pairs of potentials. 
The only correlator that must be calculated is the 2-point correlator of pre-collision potentials from the same ion. We define
\bea
&& \delta_{ab} B_1^{ij}(\vec{x}_\perp,\vec y_\perp) \equiv  \lim_{{\rm w} \to 0} \langle \beta_{1\,a}^i(x^-,\vec x_\perp) \beta_{1\,b}^j(y^-,\vec y_\perp)\rangle \nn\\  
&& \delta_{ab} B_2^{ij}(\vec{x}_\perp,\vec y_\perp) \equiv  \lim_{{\rm w} \to 0} \langle \beta_{2\,a}^i(x^+,\vec x_\perp) \beta_{2\,b}^j(y^+,\vec y_\perp)\rangle
\,. \label{core5-20}
\eea
Our calculation of the functions $B_1(\vec x_\perp,\vec y_\perp)$ and $B_2(\vec x_\perp,\vec y_\perp)$ is in Appendix \ref{BBxx}. Using the index $n\in\{1,2\}$ to indicate the two ions,
the result is 
\bea
&& B_n^{ij}(\vec{x}_\perp,\vec y_\perp) = \frac{2 }{g^2 N_c \tilde\Gamma_n(\vec x_\perp,\vec y_\perp)}  \; \left({\rm exp}[\frac{g^4 N_c}{2}\;\tilde\Gamma_n(\vec x_\perp,\vec y_\perp) ]-1\right)\;
\partial_x^i \partial_y^j \tilde\gamma_n(\vec x_\perp,\vec y_\perp) \,\label{B-res}
\eea
with 
\bea
\tilde\Gamma_n(\vec x_\perp,\vec y_\perp) = 2\tilde\gamma_n(\vec x_\perp,\vec y_\perp) - \tilde\gamma_n(\vec x_\perp,\vec x_\perp) - \tilde\gamma_n(\vec y_\perp,\vec y_\perp)  \label{Gamma-tilde-def}
\eea
and
\bea
&& \tilde \gamma_n(\vec x_\perp,\vec y_\perp) = \int d^2 z_\perp \, \mu_n(\vec z_\perp)\, G(\vec x_\perp-\vec z_\perp)\,G(\vec y_\perp-\vec z_\perp) \,.
\label{gamma-tilde-def}
\eea
We note that the expressions for the correlators $B_1^{ij}(\vec{x}_\perp,\vec y_\perp)$ and $B_2^{ij}(\vec{x}_\perp,\vec y_\perp)$ will be different only if the charge distributions of the corresponding ions, $\mu_1(\vec z_\perp)$ and $\mu_2(\vec z_\perp)$, are different. 

All higher order correlators are obtained from the results in equations (\ref{core5-20}, \ref{B-res}). For example, the average of four potentials, two from each ion, is
\bea
&& \lim_{{\rm w} \to 0} \langle\beta^i_{1\,a}(x^-,\vec x_\perp)\beta^j_{1\,b}(y^-,\vec y_\perp)\,\beta^l_{2\,c}(x^+,\vec x_\perp)\beta^m_{2\,d}(y^+,\vec y_\perp) \rangle 
   = \delta_{ab}\delta_{cd}\; B^{ij}(\vec x_\perp,\vec y_\perp) \,  B^{lm}(\vec x_\perp,\vec y_\perp)\,.\nonumber \\\label{wick}
\eea 
When one of the potentials is differentiated with respect to a transverse coordinate we have, for example,
\bea
&& \lim_{{\rm w} \to 0} \langle\partial^k_x \beta^i_{1\,a}(x^-,\vec x_\perp)\beta^j_{1\,b}(y^-,\vec y_\perp)\rangle = \delta_{ab}  \partial^k_x B^{ij}(\vec x_\perp,\vec y_\perp)\,.
\eea
The average of the product of six potentials, two from the first ion and four from the second ion, is (omitting arguments)
\bea
\langle \beta_1^i \beta_1^j \, \beta_2^l \beta_2^m \beta_2^k \beta_2^r\rangle 
= 
\langle \beta_1^i \beta_1^j\rangle\,\big(\langle \beta_2^l \beta_2^m \rangle \langle\beta_2^k \beta_2^r\rangle 
+ \langle \beta_2^l \beta_2^k \rangle \langle\beta_2^m \beta_2^r\rangle 
+ \langle \beta_2^l \beta_2^r \rangle \langle\beta_2^k \beta_2^m\rangle  \big)\,.\label{example-6}
\eea
In the limit that the width w is taken to zero, each of the 2-point correlators on the right side of equation (\ref{example-6}) can be rewritten using equation (\ref{core5-20}).

We close this section with a comment about the Glasma Graph approximation. This approximation has been used in all previous near-field calculations, and we use it throughout this paper. 
However, it is important to note that our calculation in fact provides a way to test its validity. We will discuss this in section \ref{sec-glasma-graph}.

\section{Regulation and physical scales}
\label{sec-reg}

The result of the previous section is an expression for the energy-momentum tensor in terms of the 2-point correlation function $B^{ij}(\vec x_\perp,\vec y_\perp)$ which is given in equations (\ref{B-res}, \ref{Gamma-tilde-def}, \ref{gamma-tilde-def}) and (\ref{Gf-def}).
The last step in the calculation is to take the limit that the relative coordinate $\vec r=\vec x_\perp-\vec y_\perp~\to~0$. This limit produces a divergence that must be regulated. This is not unexpected since the CGC model we are using is classical and breaks down at small distances.
An infra-red regulator is also needed to define the Green's functions from which the potential correlators are constructed. 
In this section we give a general discussion of the parameters that are needed to regulate the energy-momentum tensor, and their relation to the physical scales in the problem. 

\subsection{Infra-red regulator}
\label{sec-ir-reg}
In equation (\ref{Gf-def}) we introduced an infra-red regulator which we called $m$. 
To obtain more insight into how this regulator should be chosen, we expand the Green's function in (\ref{Gf-def}) around $m=0$ which gives
\bea
\label{Gf-exp}
G(m\,r) \approx \frac{1}{2\pi}\ln\left(\frac{L}{r}\right)~~ \text{with}~~L = \frac{2 e^{-\gamma_E}}{m}
\eea
where $\gamma_E\approx 0.577$ is  Euler's constant. 
Since the valence parton sources come from individual nucleons, confinement tells us that their effects should die off at transverse length scales larger than $1/\Lambda_{\rm QCD}$, and the Green's function should therefore be defined with boundary conditions so that it vanishes at $r \gtrsim 1/\Lambda_{\rm QCD}$. From equation (\ref{Gf-exp}) we see that we should choose $m\sim \Lambda_{\rm QCD}$.

\subsection{Ultra-violet regulator}

In this sub-section we discuss how to take the limit that the relative coordinate $\vec r$ goes to zero. 
We start with a quick review of some of the notation that has already been introduced. 
In equation (\ref{big-in}) we introduced the colour charge density for the first ion $\lambda_1(x^-,\vec x_\perp)$, which enters the Green's function $\gamma_1(x^-,\vec x_\perp,\vec y_\perp)$ that is defined in equation (\ref{gamma-def-0}). 
After taking the limit that the width of the source charge density across the light-cone goes to zero (see Appendix \ref{BBxx}), these functions are effectively replaced, respectively, by $\mu_1(\vec x_\perp)$ and $\tilde\gamma_1(\vec x_\perp, \vec y_\perp)$, which are defined in equations (\ref{norm}, \ref{gamma-tilde-def}). 

The two transverse coordinates $\vec x_\perp$ and $\vec y_\perp$ can be written in terms of the relative and average coordinates as $\vec x_\perp = \vec R+\vec r/2$ and $\vec y_\perp = \vec R - \vec r/2$ (see equation (\ref{rR-def})). 
The original MV model made the simplifying assumption that the colliding nuclei are infinitely large in the transverse directions and invariant under rotations and translations, which means that the functions $\mu_1$ and $\mu_2$ are assumed constant. 
We have calculated the energy-momentum tensor using non-constant colour density functions, and we study in detail the effects of these functions in our companion paper \cite{carrington1}. 
In this paper we present results only for the simpler case of homogeneous nuclei, and we write  $\mu_1(\vec x_\perp) = \mu_2(\vec x_\perp) \equiv\bar\mu$.
In this approximation the Green's functions $\tilde\gamma_1(\vec x_\perp, \vec y_\perp)$ and $\tilde\gamma_2(\vec x_\perp, \vec y_\perp)$ are equal to each other and depend only on the magnitude of the relative coordinate $r = |\vec x_\perp - \vec y_\perp|$, and will  both be denoted $\bar\gamma(r)$. 
Using this notation equation (\ref{B-res}) becomes
\bea
B^{ij}(\vec x_\perp,\vec y_\perp) = g^2\,\left(\frac{e^{g^4 N_c \left(\bar\gamma(r)-\bar\gamma(0)\right)}-1}{g^4 N_c \left(\bar\gamma(r)-\bar\gamma(0)\right)} \right)\partial_x^i \partial_y^j \bar\gamma(r) \label{B-constant}
\eea
with
\bea
\bar \gamma(r)  = \frac{\bar\mu}{4 \pi  m} r K_1(m r) \,. 
\,\label{gamma-k-const}
\eea
Substituting (\ref{gamma-k-const}) into (\ref{B-constant}) one sees immediately that the correlator $B^{ij}$ diverges logarithmically when $r\to 0$. 

One approach to regulate this divergence, and the divergences that appear in derivatives of the 2-point correlator, is to expand in $r$ and regulate any factors involving inverse powers of $r$, or logarithms of $r$, by making the replacement $r\to 1/Q_s$.  
In this paper we use a different method  \cite{Fujii:2008km} which is more suitable for a calculation of the energy-momentum tensor, where we want to take the relative coordinate $r$ strictly to zero. We rewrite the function $\bar\gamma(r)$ as a momentum integral 
\bea
\bar \gamma(r)  = \frac{\bar\mu}{4 \pi  m} r K_1(m r) = \bar \mu \int \frac{d^2k}{(2\pi)^2} \frac{e^{i\vec r\cdot \vec k} }{\left(k^2+m^2\right)^2}\,
\,\label{gamma-k-const-2}
\eea
and then substituting (\ref{gamma-k-const-2}) into (\ref{B-constant}) we  obtain 
\bea
\lim_{r\to 0}B^{ij}(\vec x_\perp,\vec y_\perp) = g^2 \bar \mu \int \frac{d^2k}{(2\pi)^2} \frac{\hat k^i \hat k^j k^3}{\left(k^2+m^2\right)^2} = \delta^{ij}g^2\frac{\bar \mu}{4\pi} \int_0^\infty dk\, \frac{k^3}{\left(k^2+m^2\right)^2}\,.\label{log-div}
\eea
In the last line we have used  $\hat k^i \hat k^j\to \delta^{ij}/2$ which follows from the integration over the angular variables. 
We introduce a momentum cutoff which we call $\Lambda$ to regulate the logarithmically divergent integral 
in equation (\ref{log-div}), which gives 
\bea
\lim_{r\to 0}B^{ij}(\vec x_\perp,\vec y_\perp) =\delta^{ij}g^2\frac{\bar \mu}{8\pi}\left({\rm LN}-\frac{\Lambda^2}{\Lambda^2+m^2}\right)\,
\label{corr-res-basic}
\eea
where we have defined 
\bea
{\rm LN} = \ln\left(\frac{\Lambda^2}{m^2}+1\right)\,.
\eea
%We comment that the notation $\Lambda$ was also used to denote the covariant gauge pre-collision potential in equation (\ref{resid-1b}), but there is no overlap between sections that reference these two variables. 

The derivatives of correlators can be calculated the same way. 
After integrating over the angular variables, products of odd numbers of unit vectors $\hat k$ give zero and products of four and six unit vectors are replaced with sums of products of delta functions using
\bea
&& \hat k^i \hat k^j \hat k^k \hat k^l \to \frac{1}{8}\left( \delta_{il} \delta_{jk}+\delta_{ik} \delta_{jl}+\delta_{ij} \delta_{kl} \right) \nonumber \\
&& \hat k^i \hat k^j \hat k^k \hat k^l \hat k^m \hat k^n \to \frac{1}{48}\left(\right.\delta _{in} \delta _{jm} \delta _{kl}+\delta _{im} \delta _{jn} \delta
   _{kl}+\delta _{ij} \delta _{kl} \delta _{mn} 
   +\delta _{in} \delta _{jl} \delta
   _{km}+\delta _{il} \delta _{jn} \delta _{km} \nonumber \\
&& ~~~~~~~~~~~~~~~~  +\delta _{im} \delta _{jl} \delta
   _{kn}  +\delta _{il} \delta _{jm} \delta _{kn}+\delta _{in} \delta _{jk} \delta
   _{lm}+\delta _{ik} \delta _{jn} \delta _{lm}+\delta _{ij} \delta _{kn} \delta
   _{lm}+\delta _{im} \delta _{jk} \delta _{ln} \nonumber \\
   &&~~~~~~~~~~~~~~~~ +\delta _{ik} \delta _{jm} \delta
   _{ln}+\delta _{ij} \delta _{km} \delta _{ln}+\delta _{il} \delta _{jk} \delta
   _{mn}+\delta _{ik} \delta _{jl} \delta _{mn}\left.\right)\,.\nonumber
\eea
When two derivatives act on the 2-point correlator we obtain 
\bea
&& \lim_{r\to 0} \partial^l_{n_1} \partial^m_{n_2} B^{ij}(\vec{x}_\perp, \vec{y}_\perp) \label{alina-116} \\
&=&
(-1)^{n_1+n_2+1} \frac{g^2 \bar \mu }{16\pi} 
\bigg[(\delta^{ij}\delta^{lm} + \delta^{il}\delta^{jm} + \delta^{im}\delta^{jl})
\Big( \frac{\Lambda^4}{2(\Lambda^2+m^2)} -m^2 {\rm LN}  +\frac{\Lambda^2 m^2}{\Lambda^2+m^2} \Big) \nn\\
&& + \delta^{lm}\delta^{ij}  \frac{3g^4 \bar \mu}{8\pi} 
\left({\rm LN}^2 + \frac{\Lambda^4}{(\Lambda^2+m^2)^2} - \frac{2{\rm LN}\Lambda^2}{\Lambda^2+m^2}\right) \bigg] \nn
\eea
where we use the notation that the index $n_i$ is 1 if the transverse derivatives are with respect to $\vec x_\perp$ and 2 if the transverse derivatives are with respect to $\vec y_\perp$. We note that the leading order contributions to equation (\ref{alina-116}) should agree with equation (27) in Ref. \cite{Fujii:2008km}, but we have a different sign for the last term in that equation.

We comment that although one might expect that the alternative regularization scheme described above equation (\ref{gamma-k-const-2})
 would give very similar results, this is not always true. For some correlators the only difference is a redefinition of the mass scale which appears in the argument of a logarithmic factor, but in some cases the two regularization methods give parametrically different results. 
We consider as an example the biggest contribution to (\ref{alina-116}) with $n_1=1$ and $n_2=2$ which is 
\bea
\left[\partial^l_x \partial^m_y B^{ij}(\vec x_\perp,\vec y_\perp)\right]_{\rm lo} = \frac{g^2 \bar\mu \Lambda^2}{32\pi}
\big(\delta^{ij}\delta^{lm} + \delta^{il}\delta^{jm} + \delta^{im}\delta^{jl}  \big)\,. \label{116-lo}
\eea
If we start from equation (\ref{B-constant}), take the derivatives, expand in $r$, and regulate divergences using $r\to 1/Q_s$, the leading order contribution is 
\bea
\left[\partial^l_x \partial^m_y B^{ij}(\vec x_\perp,\vec y_\perp)\right]_{\rm alternate} = \frac{g^2\bar\mu  m^2 LG}{16\pi}\big(\delta^{ij}\delta^{lm} + \delta^{il}\delta^{jm} + \delta^{im}\delta^{jl}  \big) \label{116-no}
\eea
where we have introduced the definitions
\bea
 LG \equiv \ln[(\hat m /Q_s)^2]  \text{~~and~~} \hat m \equiv \frac{m}{2}e^{\gamma_E+\frac{1}{2}}
  \,.\label{mhat-def}
\eea
Comparing equations (\ref{116-lo}, \ref{116-no}) we see that when we write the correlator as a momentum integral with an ultra-violet cutoff, the regulated result is $\sim \Lambda^2$, but if we expand in $r$ and make the replacement $r\to 1/Q_s$ the regulated expression is $\sim m^2$. The expansion method would therefore replace the ultra-violet regulator with the infra-red regulator. 

In Ref. \cite{Chen:2015wia} both of these regularization methods are used, depending on which correlator is being calculated. 
In our calculation we consistently regularize by writing each correlator as a momentum integral, differentiating as needed, taking the limit $r\to 0$, and calculating the regulated momentum integral. We note also that we treat both the 1-point correlator $B^{ij}(\vec x_\perp,\vec x_\perp)$ and the 2-point correlator $B^{ij}(\vec x_\perp,\vec y_\perp)$ in the same way. 

\subsection{The MV scale}
\label{sec-MV-scale}

The charge density provides another dimensionful scale which is usually called the MV scale and defined in our notation as $g^2 \sqrt{\bar\mu}$. 
This scale is related to the saturation scale $Q_s$, although the exact relationship between them cannot be determined within the CGC approach, as discussed in many papers (see for example \cite{Iancu:2003xm, Lappi:2007ku}). 
Expanding in $mr\ll 1$ equation (\ref{Gamma-tilde-def}) gives
\bea
[\tilde\Gamma(\vec x_\perp,\vec y_\perp)]_{\rm MV} = \bar \mu\frac{r^2}{8\pi}
\left(2\ln\left(\frac{mr}{2}\right)+2\gamma_E -1  + {\cal O}\left((mr)^2\right) \right) \,\label{MVc}
\eea
and using (\ref{mhat-def}) we rewrite equation (\ref{MVc}) as 
\bea
[\tilde\Gamma(\vec x_\perp,\vec y_\perp)]_{\rm MV} = \frac{\bar \mu r^2}{8\pi}
\left(LG-2  + {\cal O}\left((\hat m\, r)^2\right) \right)\,.\label{MVd}
\eea
Substituting (\ref{MVc}) into (\ref{B-res}) we obtain 
\bea
[B^{ij}(\vec x_\perp,\vec y_\perp)]_{\rm MV}  \approx \frac{2}{g^2 N_c r^2}\delta^{ij}  \big[1-\delta^{\epsilon}\big]\,\label{MVh}
\eea
where we have defined two dimensionless parameters
\bea
&& \delta = \hat m r \ll1 \text{~~ and ~~}  \epsilon = \frac{ g^4 N_c}{8\pi} \bar\mu r^2\,.
\eea

The result in (\ref{MVh}) can be used to argue that the MV scale is proportional to the saturation scale \cite{JalilianMarian:1996xn}.
For $\epsilon<1$ we have $1-\delta^{\epsilon} \approx -\epsilon\ln\delta$ 
and (\ref{MVh}) becomes
\bea
\label{Ai0-Aj0-approx-1}
[B^{ij}(\vec x_\perp,\vec y_\perp)]_{\rm MV}     
\approx -\frac{g^2\bar \mu}{8 \pi}   \delta^{ij}  \ln (\hat m^2 r^2) \,.
\eea
For $\epsilon>1$ we have $1-\delta^{\epsilon} \approx 1$ 
and (\ref{MVh}) takes the form
\bea
\label{Ai0-Aj0-approx-2}
[B^{ij}(\vec x_\perp,\vec y_\perp)]_{\rm MV}     
\approx \frac{2 \delta^{ij} }{g^2 N_c} \frac{1}{r^2}  .
\eea
The condition $\epsilon<1$ corresponds to small transverse distances, or momentum scales $k_\perp \gtrsim \bar\alpha \sqrt{\bar\mu}$ with $\bar\alpha = g^2\sqrt{N_c/(8\pi)}$.
The opposite case $\epsilon>1$ means transverse distances that are large (but still much less than $1/\Lambda_{\rm QCD}$), or momentum scales that satisfy $\bar\alpha \sqrt{\bar \mu} > k_\perp > \Lambda_{\rm QCD}$.
From the Fourier transforms of equations (\ref{Ai0-Aj0-approx-1}, \ref{Ai0-Aj0-approx-2}) we see that at small transverse distances, or large momentum scales, the correlation function falls like $1/k_\perp^2$ (perturbative behaviour) and at large transverse distances, or small momentum scales, the correlation function rises like $\sim \ln(k_\perp)$. 
The number of gluons in a range $dk_\perp$ about some $k_\perp$ is related to the trace of the gluon propagator in equation (\ref{MVh}),  multiplied by the phase space factor $k_\perp$.
The peak occurs approximately at the momentum scale that corresponds to $\epsilon=1$, which divides the growing and falling regions of the distribution function. 
Therefore the typical transverse momenta of gluons, or the saturation scale, satisfies
\bea
\epsilon\big|_{r\sim 1/Q_s}  = 2\pi \alpha_s^2 N_c r^2\bar \mu\big|_{r \sim 1/Q_s}  \equiv 1 \,
\label{A-def}
\eea 
which gives $Q_s^2 \sim g^4\bar\mu$. 
As mentioned above, the proportionality factor cannot be determined within the CGC approach. We define
\bea
Q_s^2 = g^4  \bar\mu\,. \label{myUbar}
\eea
In the next section we obtain numerical results for the energy density and transverse and longitudinal pressures using this definition. 
The ratios of these quantities as functions of $\tau$ give information about the equilibration of the system,  but the numerical values of the energy and pressures themselves should only be considered order-of-magnitude estimates.

\section{Results and discussion}
\label{sec-results}

\iffalse
\subsection{Consistency checks}
\label{sec-cc}

After all correlators have been calculated, we obtain a final result for the energy-momentum tensor that depends on the saturation scale $Q_s$, the ultra-violet cutoff $\Lambda$, the infra-red regulator $m$, the proper time $\tau$ (which is our expansion parameter), the strong coupling constant $g$. 
%
We have checked that our result for the energy-momentum tensor is traceless, symmetric and divergenceless. 
%We have also checked that the divergence of the energy-momentum tensor is zero, by verifying that $\partial_\mu^{\rm mink}T^{\mu\nu}_{\rm mink}(\tau,\eta,\vec R)=0$, where the derivative operator is given in equation (\ref{gradient}). 

\fi
\subsection{Analytic results}
\label{sec-res-ana}

%In this section we present some analytic results. 
The  energy-momentum tensor has the form
\bea
T = \left(
\begin{array}{cccc}
 T^{00} & T^{01} & 0 & 0 \\
 T^{10} & T^{11} & 0 & 0 \\
 0 & 0 & T^{22} & 0 \\
 0 & 0 & 0 & T^{33} \\
\end{array}
\right)\,.
\label{TT-fake}
\eea
To fourth order in $\tau$ the non-zero components can be written
\bea
&& T^{00} = {\cal E}_0 + (2-\cosh(2\eta)) {\cal E}_2\,\tau^2 +  (3-2\cosh(2\eta)) {\cal E}_4 \tau^4 \nn\\
-&&T^{11} = {\cal E}_0 + (2+\cosh(2\eta)) {\cal E}_2\,\tau^2 +  (3+2\cosh(2\eta)) {\cal E}_4 \tau^4 \nn\\
&&T^{22}=T^{33} = {\cal E}_0 + 2 {\cal E}_2\,\tau^2 +  3 {\cal E}_4 \tau^4 \nn \\
&& T^{01}=T^{10} =   \cosh(\eta)\,\sinh(\eta) C_2 \tau^2 + \cosh(\eta)\,\sinh(\eta) C_4 \tau^4 \,.
\label{ana1}
\eea
For simplicity we give our analytic results for the coefficients in equation (\ref{ana1}) by expanding in $m/\Lambda$ and including $m$ dependence only in the arguments of the logarithms. We have
\bea
\tilde{\cal E}_0 &\equiv& \frac{3 g^6 \bar\mu^2 \ln ^2\left(\Lambda/m\right)}{\pi ^2} \nn\\
{\cal E}_0 &=& \tilde{\cal E}_0 \left(1- \frac{1}{2\ln(\Lambda/m)} \right)^2  \nn \\[2mm]
 {\cal E}_2 &=& - \tilde{\cal E}_0 \bigg[
\frac{\Lambda^2}{4 \ln \left(\Lambda/m\right)}\left(1-\frac{1}{2\ln\left(\Lambda/m\right) }\right)
+\frac{3 g^4 \bar\mu \ln(\Lambda/m)}{\pi}\left(1-\frac{1}{2\ln(\Lambda/m)}\right)^3 
\bigg] \nn\\[4mm]
{\cal E}_4 &=& \tilde{\cal E}_0 
\bigg[
\frac{\Lambda^4}{64 \ln(\Lambda/m)}\left(1+\frac{3}{2\ln(\Lambda/m)}\right)
 + \frac{183 g^4 \bar\mu \Lambda^2}{128\pi}\left(1 -\frac{1}{2\ln(\Lambda/m)}\right)^2 \nn\\
&& ~~~~ +\frac{13767 g^8 \bar\mu^2 \ln^2(\Lambda/m)}{2048 \pi^2}\left(1-\frac{1}{2\ln(\Lambda/m)}\right)^4
\bigg]   \nn \\[4mm]
 C_2 &=& \tilde{\cal E}_0
\bigg[\frac{\Lambda^2}{2 \ln \left(\Lambda /m\right)}\left(1-\frac{1}{2\ln(\Lambda/m)}\right)
+\frac{6g^4\bar\mu \ln(\Lambda/m)}{\pi}\left(1-\frac{1}{2\ln(\Lambda/m)}\right)^3 \nn
\bigg]
 \nn
 \eea
 \bea
\hspace*{-.4cm} C_4 &=&  \tilde{\cal E}_0 
\bigg[
\frac{\Lambda^4}{16\ln(\Lambda/m)}\left(1+\frac{3}{2\ln(\Lambda/m)}\right)
+ \frac{183g^4\bar\mu\Lambda^2}{32\pi}\left(1-\frac{1}{2\ln(\Lambda/m)}\right)^2 \nn\\
\hspace*{-.4cm} && ~~~~ +\frac{13767 g^8 \bar\mu^2 \ln^2(\Lambda/m)}{512\pi^2}\left(1-\frac{1}{2\ln(\Lambda/m)}\right)^4 
\bigg]\,. \label{ana2}
\eea
At sixth order the expressions are very lengthy and will not be given explicitly.

\subsection{Accuracy of the  Glasma Graph approximation}
\label{sec-glasma-graph}

Throughout this paper we have used the Glasma Graph approximation which allows us to use Wick's theorem to calculate products of light-cone gauge potentials.  
In spite of the fact that we have used this approximation consistently throughout, we can still obtain a quantitative measure of its validity in our calculation.

The initial longitudinal magnetic field in equation (\ref{Bz-initial}) is rewritten using the initial condition (\ref{cond2}) as 
\bea
B = B^z(0,\vec x_\perp) =  F_1^{21}(\vec x_\perp) + F_2^{21}(\vec x_\perp) +i g \epsilon^{ij}[\beta_1^i(\vec x_\perp),\beta_2^j(\vec x_\perp)]\,.
\label{B-long}
\eea
If we use the fact that the pre-collision potentials are pure gauge (see equations (\ref{pg1}, \ref{pg2})),  equation (\ref{B-long}) can be written as 
\bea
B_{\rm pg} =  i g \epsilon^{ij}[\beta_1^i(\vec x_\perp),\beta_2^j(\vec x_\perp)]\,.
\label{B-shorty}
\eea
We should find that our averaging procedure gives $\langle B\rangle = \langle B_{\rm pg}\rangle$ or $\langle F^{21}_1\rangle = \langle F^{21}_2\rangle= \langle F^{12}_1\rangle = \langle F^{12}_2\rangle =0$. We consider  the field-strength from the first ion, and suppress the subscript 1. 
Expanding out all terms we have
\bea
F^{12}(\vec x_\perp) = \partial_x^1 \beta^2(\vec x_\perp) - \partial_x^2 \beta^1(\vec x_\perp)  - i g \beta^1(\vec x_\perp)\beta^2(\vec x_\perp) + i g\beta^2(\vec x_\perp)\beta^1(\vec x_\perp) \,.
\eea 
The expectation value of this expression is obviously zero because the averaging procedure
involves a trace over colour indices.
In fact it is zero even before the colour trace is taken, since
the first two terms have only one potential (all terms with an odd number of potentials are set to zero), and the third and fourth term give zero using equation (\ref{corr-res-basic}). 

Next we consider the expectation value
\bea
\langle F^{12}(\vec x_\perp) \partial^2_y \beta^1(\vec y_\perp)\, \rangle\,.
\eea
This should also give zero, since we have $F^{12}(\vec x_\perp)=0$ directly from equation (\ref{pg1}). The terms with three potentials are set to zero, and the terms with two potentials give
\bea
\lim_{r\to 0}\langle F^{12}(\vec x_\perp) \partial^2_y \beta^1(\vec y_\perp)\, \rangle  = \lim_{r\to 0} \frac{1}{2}(N_c^2-1) \left(\partial^1_x\partial^2_y B^{21}(\vec x_\perp,\vec y_\perp) - \partial^2_x\partial^2_y B^{11}(\vec x_\perp,\vec y_\perp)\right)\,.
\label{pg-bad}
\eea
Equation (\ref{alina-116}) gives that the right side of (\ref{pg-bad}) is not zero. 
This apparent contradiction is related to the Glasma Graph approximation, which sets all correlators with odd numbers of potentials to zero. 

In equation (\ref{comps}) we gave expressions for the field components to order $\tau^4$ in terms of the lowest order magnetic field. We have just shown that, using the Glasma Graph approximation, the result for the energy-momentum tensor will be different depending on whether we use the expression for $B$ in equation (\ref{B-long}) or equation (\ref{B-shorty}). We can therefore test the approximation by comparing the final expressions for the energy-momentum tensor obtained from the two different forms of the lowest order magnetic field. Some numerical results from this comparison are presented in the next section. 
We note that the results in equations (\ref{ana1}, \ref{ana2}) were obtained using equation (\ref{B-long}).

\subsection{Numerical results}

The units for energy and pressure are GeV  and for lengths and times we use fm (we use natural units $\hbar = c=1$).
Unless stated otherwise, we use  $g=1$, $N_c=3$, $m=0.2$ GeV and $\Lambda = Q_s=2$ GeV.
We use always the expressions for the energy-momentum tensor obtained using equation (\ref{B-long}), except when we explicitly say otherwise. 

\begin{figure}[htb]
\centering
\includegraphics[scale=0.8]{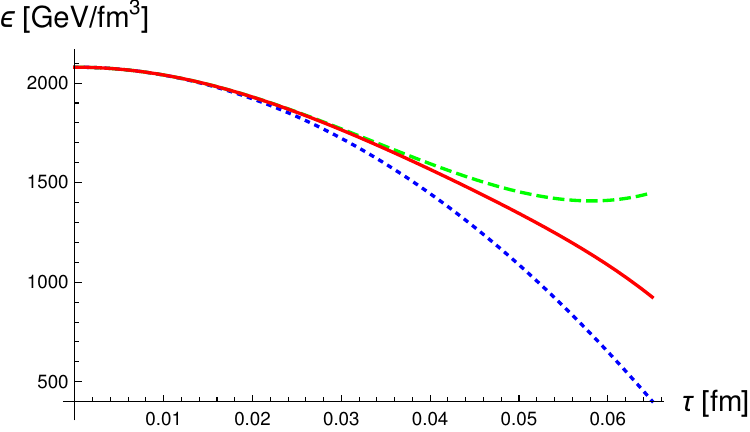}
\caption{The energy density as a function of $\tau$ at $\eta=0$.  The blue (dotted), green (dashed) and red  lines are, respectively, the results to order $\tau^2$, $\tau^4$  and  $\tau^6$. 
\label{energy-plot}}
\end{figure}

In figure \ref{energy-plot} we show the energy density as a function of $\tau$. 
The initial energy density is ${\cal E}_0$ = 2080 GeV/fm$^3$ and we note that although this number is only an order-of-magnitude approximation, as discussed at the end of section \ref{sec-MV-scale}, it is not far from the estimate from Ref. \cite{Mrowczynski:2017kso} for the fraction of the collision
energy that goes into particle production.

Next we want to study the possible equilibration of the system. 
At $\tau=0^+$ the energy-momentum tensor has the diagonal form 
\bea
T_{\rm mink}^{\rm initial} = \left(
\begin{array}{cccc}
{\cal E}_0  & 0 & 0 & 0 \\
0 & - {\cal E}_0  & 0 & 0  \\
0 & 0 & {\cal E}_0  & 0  \\
0 & 0 & 0 & {\cal E}_0  
\end{array}
\right)\,.\label{diag}
\eea
The longitudinal pressure is large and negative and the system is far from equilibrium. 
We look at the evolution of the energy density and pressure as functions of time. 
We define the normalized longitudinal and transverse pressures as
\bea
\frac{p_L}{{\cal E}} = \frac{T_{\rm mink}^{11}}{T_{\rm mink}^{00}} \text{~~ and ~~} \frac{p_T}{{\cal E}} = \frac{1}{2}\frac{(T_{\rm mink}^{22} + T_{\rm mink}^{33})}{T_{\rm mink}^{00}}\,.
\eea
If the system approaches equilibrium, the longitudinal pressure must grow as the system evolves. 
The energy-momentum tensor is traceless at all times ($T^\mu_{~\mu}=0$) and therefore the normalized transverse pressure must decrease as the normalized longitudinal pressure increases. 

\begin{figure}[htb]
\centering
\includegraphics[scale=0.8]{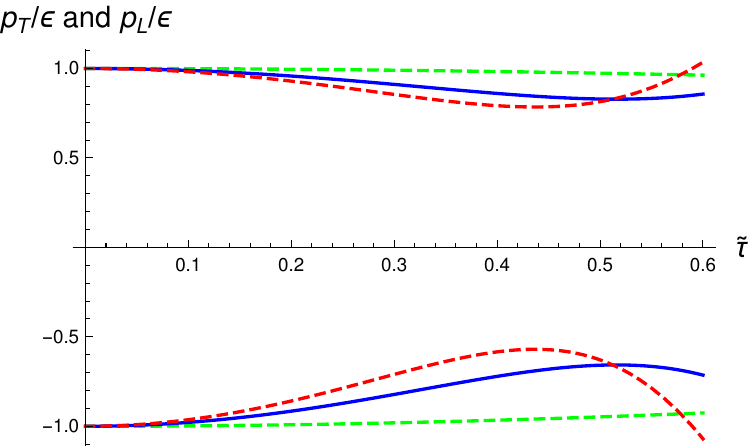}
\caption{The normalized longitudinal and transverse pressures to order $\tau^4$ versus $\tilde\tau$ at $\eta=0$. 
The blue (solid) line shows the result obtained using equation (\ref{B-shorty}) and the red (dashed) line is the result from equation (\ref{B-long}). The green (dotted) lines are the leading order approximation.
In each case the lower line is $p_L/{\cal E}$ and the upper line is $p_T/{\cal E}$. 
\label{isot}}
\end{figure}

\begin{figure}[H]
\centering
\includegraphics[scale=0.8]{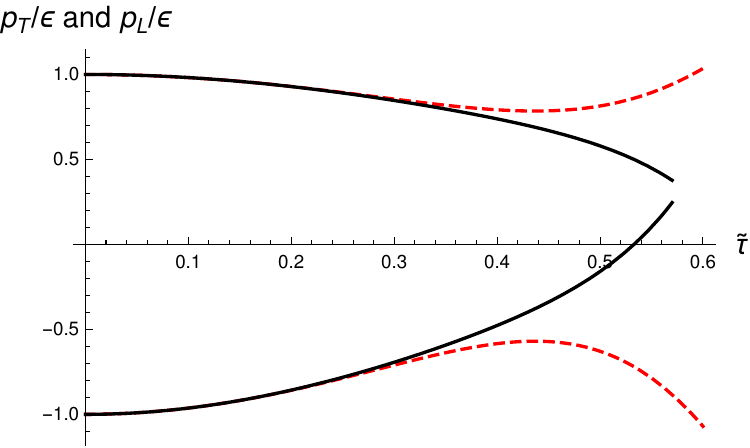}
\caption{The normalized longitudinal and transverse pressures versus $\tilde\tau$ at $\eta=0$. 
The red (dashed) and black (solid) lines are respectively the results to order $\tau^4$ and $\tau^6$. 
The lower lines are  $p_L/{\cal E}$ and the upper lines are $p_T/{\cal E}$. 
\label{isot6}}
\end{figure}

In figure \ref{isot} we show the normalized longitudinal and transverse pressures to order $\tau^4$, as functions 
of the dimensionless variable $\tilde\tau = Q_s \tau$. The red (dashed) and blue (solid) lines are the results obtained using, respectively, equations (\ref{B-long}) and (\ref{B-shorty}), and the closeness of these results is an indication of the validity of the Glasma Graph approximation. 
One sees that  the system starts to equilibrate, but the normalized pressures move apart again at $\tilde\tau\sim 0.5$, which is consistent with the breakdown of the $\tau$ expansion observed in figure \ref{energy-plot}. 
We can also compare our fourth order results with those obtained in Ref. \cite{Chen:2015wia} where the authors included only terms that they identify as leading order in $\Lambda$.  
The green (dotted) line shows this leading order result to fourth order in $\tau$.\footnote{We are able to obtain the leading order analytic results of Ref. \cite{Chen:2015wia} but we  cannot  reproduce their Fig. 5.} 
In figure \ref{isot6} we show the normalized longitudinal and transverse pressures to order $\tau^4$ and $\tau^6$. One sees that the expansion breaks down at later times when terms to order $\tau^6$ are included, and the system moves closer to the isotropic state before the breakdown occurs. 

\begin{figure}[htb]
\centering
\includegraphics[scale=0.40]{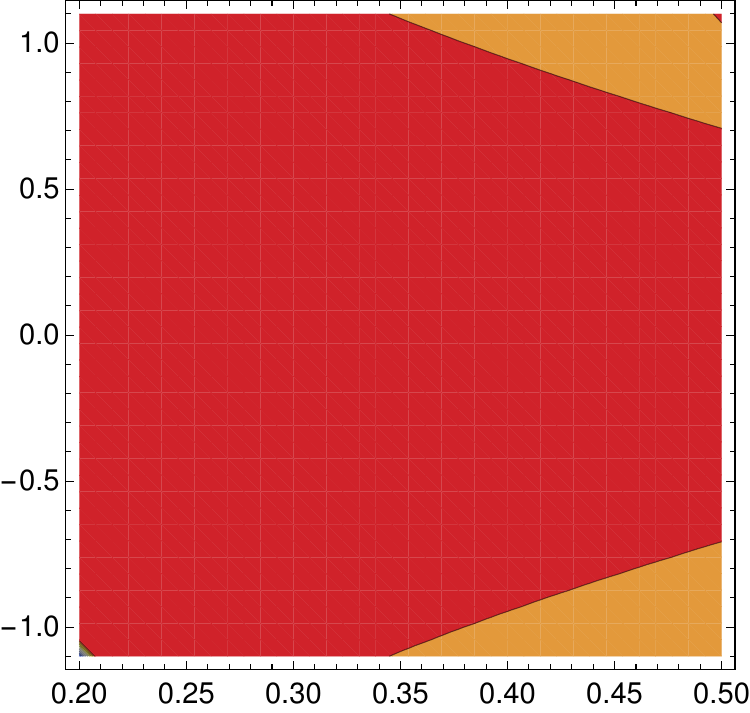}
\includegraphics[scale=0.40]{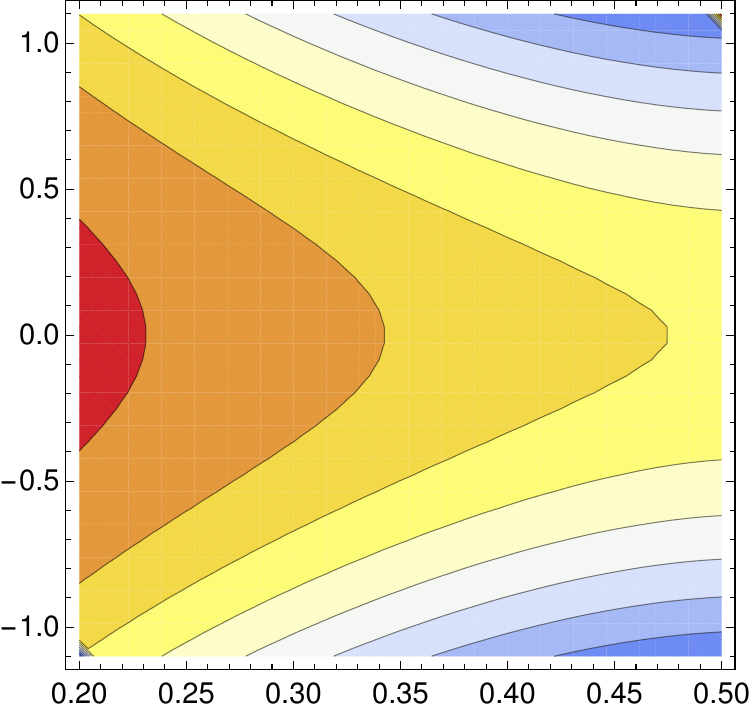}
\includegraphics[scale=0.55]{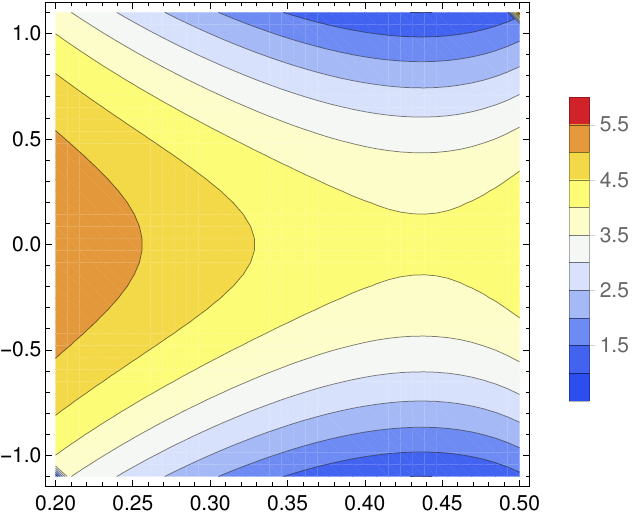}
\caption{The quantity $A_{TL}$ in equation (\ref{AA-def}) at order $\tau^4$. The vertical axis shows $\eta$ and the horizontal axis is $\tilde\tau$. The left panel shows the leading order approximation of Ref. \cite{Chen:2015wia}, the centre panel shows the result obtained using equation (\ref{B-shorty}), and the right panel is the result using equation (\ref{B-long}).
\label{plotA-one}}
\end{figure}

\begin{figure}[htb]
\centering
\includegraphics[scale=0.44]{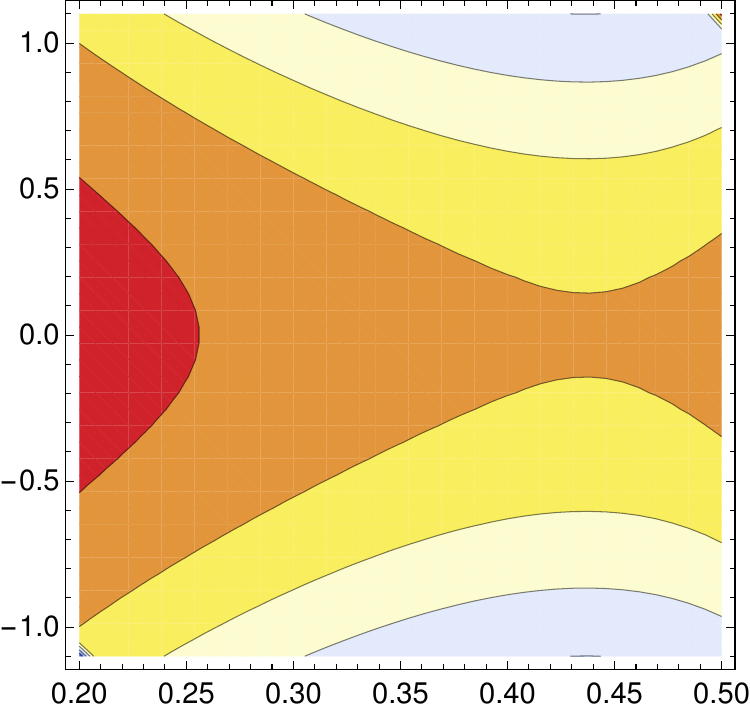}
\includegraphics[scale=0.61]{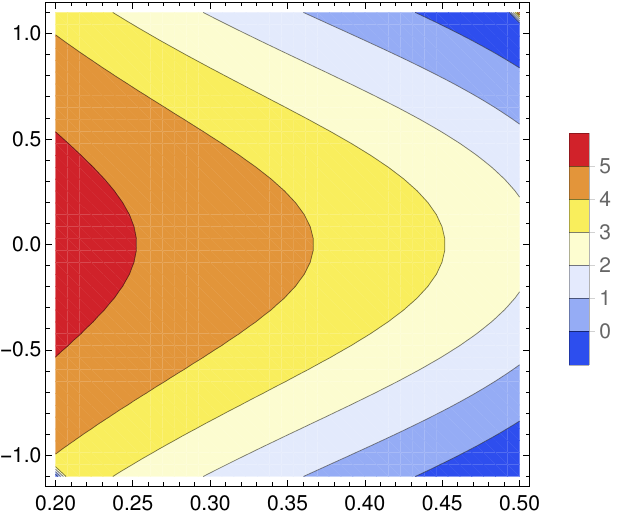}
\caption{The quantity $A_{TL}$ in equation (\ref{AA-def}) at order $\tau^4$ (left panel) and order $\tau^6$ (right panel). The vertical axis shows $\eta$ and the horizontal axis is $\tilde\tau$. 
\label{plotA-two}}
\end{figure}

The authors of Ref. \cite{Jankowski:2020itt} suggest that the evolution of the glasma can best be studied using the quantity
\bea
A_{TL} \equiv \frac{3(p_T-p_L)}{2p_T+p_L}\,\label{AA-def}
\eea
which takes the value $A_{TL}=6$ at $\tau=0$ (using equation (\ref{diag})) and would be zero in an equilibrated plasma. 
In figure \ref{plotA-one} we show $A_{TL}$ as a function of $\tilde\tau$ and $\eta$ at order $\tau^4$ using the leading order approximation of Ref. \cite{Chen:2015wia}, and without approximation but using the two different results for the magnetic field in equations (\ref{B-long}) and (\ref{B-shorty}).  One sees that the results obtained using equations (\ref{B-long}) and (\ref{B-shorty}) are fairly close to each other, but the leading order approximation does not agree well. We note that the appearance of the saddle structure in the right panel indicates the breakdown of the near field expansion. 
In figure \ref{plotA-two} we show $A_{TL}$ at order $\tau^4$ and $\tau^6$.
In the left panel we see, from the appearance of the saddle, that the fourth order calculation breaks down at $\tilde\tau\sim 0.4$. 
The right panel shows clearly that when sixth order terms are included the region for which the expansion is valid is extended, and the system evolves significantly closer to the equilibrium state.  

Our calculation is based on a classical description and we can estimate the regime of validity of this description by looking at the constraint imposed by the uncertainty principle. The classical description requires $\Delta E \Delta t \gg 1$. 
Since the energy released in the collision is extremely large, as seen in figure \ref{energy-plot}, the lower bound for the range of times that satisfy the constraint  will be very small, which is the idea that justifies the near field expansion. 
To obtain a quantitative approximation for this lower bound we estimate the initial energy as $\Delta E = {\cal E}_0 S \Delta t$ where ${\cal E}_0\approx 2000$ GeV/fm$^3$ is the initial energy density, and $S\approx 150$ ${\rm fm}^2$ is the transverse area of overlap of the colliding nuclei. 
From these numbers we obtain $\Delta t \gg 1/\sqrt{{\cal E}_0 S} \sim 8 \times 10^{-4}$ fm. 
From figures \ref{energy-plot}, \ref{isot} and \ref{isot6} we estimate that in our calculation the $\tau$ expansion breaks down for values $\tau \gtrsim 0.05$ fm.
We see therefore that the region of validity of the near field expansion reaches far beyond the lower bound at which we no longer trust the classical description we are using.

\begin{figure}[htb]
\centering
\includegraphics[scale=0.8]{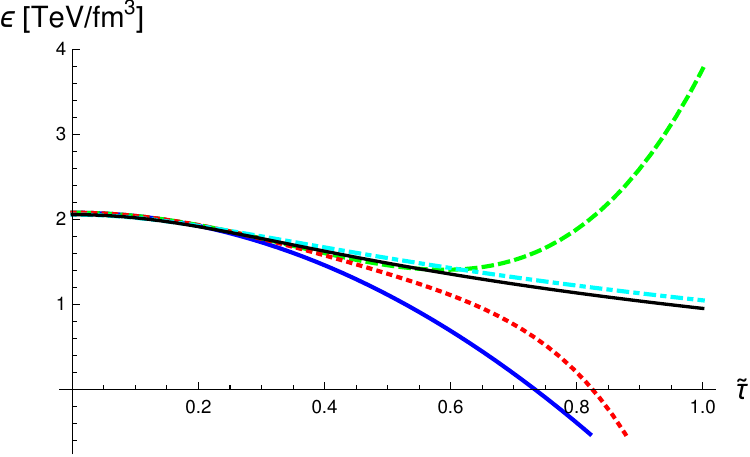}
\caption{The energy density versus $\tilde\tau$. 
The blue (solid), green (dashed) and red (dotted) lines show the results at order $\tau^2$, $\tau^4$, and $\tau^6$.
The cyan (dot-dashed) line is ${\cal E}_{\rm fit}^{(2)}$ and the thin black line just below the cyan line is ${\cal E}_{\rm fit}^{(4)}$.
The curve corresponding to ${\cal E}_{\rm fit}^{(6)}$ is directly on top of the black line and is not shown. 
\label{fuk-resum}}
\end{figure}

In Ref. \cite{Fukushima:2007yk} a different method was suggested to extract physics from the $\tau$ expanded expression for the energy. 
The authors propose that the ultra-violet regulator $\Lambda$ should be considered an unphysical scale that is related to the inverse lattice spacing in a numerical calculation.
In the limit that the lattice spacing goes to zero, lattice calculations show that the energy density is $\sim [\ln(\tau)]^2$ \cite{Lappi:2006hq}. 
We can recover these features from our results by fitting the energy to a function with the appropriate form. 
We denote the energy to order $\tau^n$ as ${\cal E}^{(n)}$ and define the corresponding fitted function
\bea
{\cal E}^{(n)}_{\rm fit} =  A\left[\ln\left(\frac{\Lambda^n}{m^n(1+\sum_{i=1}^{n/2} \tau^{2i}{\cal O}^{(n)}_{2i})}\right)\right]^2 \label{fuk-fit} 
\eea
where the symbol ${\cal O}^{(n)}_{2i}$ represents a polynomial of order $2i$ in the three  scales $\Lambda$, $Q_s$ and $m$.
For example, with $n=2$, the only term in the sum is 
$
{\cal O}^{(2)}_2 = c_1 \Lambda ^2 + c_2 \Lambda Q_s + c_3 Q_s^2 + + c_4 \Lambda m   + c_5 Q_s m  +  c_6 m^2
$,
where the coefficients $(c_1,c_2,c_3 \dots c_6)$ are determined as described below. 
When $n=4$ there are two terms in the sum, ${\cal O}^{(4)}_{2}$ (which has the same form as ${\cal O}^{(2)}_{2}$  with different coefficients), and ${\cal O}^{(4)}_{4}= c_7 \Lambda ^4+ c_8 \Lambda ^3 Q_s +\dots m^4$.
The function (\ref{fuk-fit}) satisfies 
$\lim_{\Lambda\to \infty}{\cal E}^{(n)}_{\rm fit} \sim (\ln(\tau))^2$, as desired. 
The coefficients (denoted by the $c$'s) are chosen so that the function matches our analytic result when it is expanded in $\tau$. 
In figure \ref{fuk-resum} we show the results of this fitting procedure. 
The graph demonstrates the rapid convergence of the expressions defined in equation (\ref{fuk-fit}), and the
convergence of the original $\tau$ expanded results to the same curve.

\section{Conclusions}
\label{sec-conc}

We have used a Colour Glass Condensate approach and obtained an analytic expression for the energy-momentum tensor to sixth order in an expansion in the proper time. 
We have shown that our calculation gives physically reasonable expressions for the energy density, and the longitudinal and transverse pressures. 

The idea of a proper time expansion, also called the ``near field expansion''  was proposed almost 15 years ago as a way to obtain analytic information about 
the properties of the plasma system in the very early stages after a heavy-ion collision. 
However, there are only a few calculations in the literature that make use of the method, 
and its convergence has never been studied. 
Our results clearly demonstrate the validity of the  near field expansion as an approach to describe the early time behaviour of heavy-ion collisions. 

Our expression for the energy-momentum tensor can also be used to obtain information about energy flow, elliptic flow, angular momentum, and other observable quantities. 
In our companion paper \cite{carrington1} we give a detailed analysis of  various experimentally relevant quantities that can be calculated from the energy-momentum tensor. 

\acknowledgments
We thank Doug Pickering for help with the numerical calculations, and 
Rainer Fries and Guangyao Chen for helpful correspondance.
This work was supported by the Natural Sciences and Engineering Research Council of Canada Discovery  Grant  program, and the National Science Centre, Poland under grant 2018/29/B/ST2/00646.

\appendix 

\section{Notation}
\label{AppendixA}
\label{apA}
%%%%%%%%%%%%%%%%%%%%%%%%%%%%%%%%%%%%%%%%%%%%%%%%%%%%%%%%%%%%%%%%%%%%%%%%%%%%%%
%%%%%%%%%%%%%%%%%%%%%%%%%%%%%%%%%%%%%%%%%%%%%%%%%%%%%%%%%%%%%%%%%%%%%%%%%%%%%%
%%%%%%%%%%%%%%%%%%%%%%%%%%%%%%%%%%%%%%%%%%%%%%%%%%%%%%%%%%%%%%%%%%%%%%%%%%%%%%
We use at different times three different coordinate systems. Minkowski, light-cone, and Milne (or co-moving) coordinates. 
The collision axis is defined to be the $z$-axis. The two transverse coordinates are always the last two elements in the position 4-vector and will be denoted $\vec x_\perp$.
We will write the position 4-vector as
\bea
&& x^\mu_{\rm mink} = (t,z,\vec x_\perp) \nn \\
&& x^\mu_{\rm lc} = (x^+,x^-,\vec x_\perp) \nn \\
&& x^\mu_{\rm milne} = (\tau,\eta,\vec x_\perp) 
\eea
with the usual definitions
\bea
&& x^+=\frac{t+z}{\sqrt{2}} \text{~~and~~} x^-=\frac{t-z}{\sqrt{2}}\\
&& \tau = \sqrt{t^2-z^2}=\sqrt{2x^+ x^-} \text{~~and~~} \eta = \frac{1}{2} \ln \left( \frac{x^+}{x^-} \right)\,.
\eea
We define the relative and average transverse coordinates
\bea
&& \vec r = \vec x_\perp-\vec y_\perp \text{~~and~~}
 \vec R = \frac{1}{2}\left(\vec y_\perp+\vec x_\perp\right) \label{rR-def}\,.
\eea
We will write unit vectors as $\hat r=\vec r/|\vec r|=\vec r/r$ and $\hat R=\vec R/|\vec R|=\vec R/R$
and use standard notation for derivatives like
\bea
\partial^i_{x} \equiv   -\frac{\partial}{\partial x_\perp^i} \text{~~and~~}
\partial^i_{R} \equiv  -\frac{\partial}{\partial R^i}\,.
\eea
In light-cone coordinates we have 
\bea
\partial^{+} = \frac{\partial}{\partial x^-} \text{~~and~~} \partial^{-} = \frac{\partial}{\partial x^+}\,.
\eea
We note that the chain rule gives
\bea
&& -\partial^i_{x}  = \frac{\partial}{\partial r^i} +\frac{1}{2} \frac{\partial}{\partial R^i}  \nonumber\\
&& -\partial^i_{y} = -\frac{\partial}{\partial r^i} +\frac{1}{2} \frac{\partial}{\partial R^i} \,.\label{chain}
\eea

The metric tensors in these three coordinate systems are $g_{\rm mink} = (1,-1,-1,-1)_{\rm diag}$ and 
\bea
g_{\rm lc} =
\left(
\begin{array}{cccc}
 0 & 1 & 0 & 0 \\
 1 & 0 & 0 & 0 \\
 0 & 0 & -1 & 0 \\
 0 & 0 & 0 & -1 \\
\end{array}
\right)\,,\qquad
g_{\rm milne}=  \left(
\begin{array}{cccc}
 1 & 0 & 0 & 0 \\
 0 & -\tau^2 & 0 & 0 \\
 0 & 0 & -1 & 0 \\
 0 & 0 & 0 & -1 \\
\end{array}
\right)\,.\label{metric}
\eea
The coordinate transformations are
\bea
&& x^\mu_{\rm mink} = M^\mu_{~\nu} x^\nu_{\rm lc}\,,\qquad 
M^\mu_{~\nu} = \frac{dx^\mu_{\rm mink}}{dx^\nu_{\rm lc}} = 
\left(
\begin{array}{cccc}
 \frac{1}{\sqrt{2}} & \frac{1}{\sqrt{2}} & 0 & 0 \\
 \frac{1}{\sqrt{2}} & -\frac{1}{\sqrt{2}} & 0 & 0 \\
 0 & 0 & 1 & 0 \\
 0 & 0 & 0 & 1 \\
\end{array}
\right) \nn \\
&& x^\mu_{\rm mink} = M^\mu_{~\nu} x^\nu_{\rm milne}\,,\qquad 
M^\mu_{~\nu} = \frac{dx^\mu_{\rm mink}}{dx^\nu_{\rm milne}} = 
\left(
\begin{array}{cccc}
 \cosh (\eta ) & \tau  \sinh (\eta ) & 0 & 0 \\
 \sinh (\eta ) & \tau  \cosh (\eta ) & 0 & 0 \\
 0 & 0 & 1 & 0 \\
 0 & 0 & 0 & 1 \\
\end{array}\label{milne2mink}
\right)\,.
\eea

We define a 4-dimensional gradient operator where the transverse components are derivatives with respect to the average coordinate $\vec R$ defined in equation (\ref{rR-def}). 
We can transform this gradient operator from Milne to Minkowski coordinates by taking the inverse of the transpose of (\ref{milne2mink}).
This gives
\bea
\partial_\mu ^{\rm mink} 
= 
\left(
\begin{array}{c}
\cosh(\eta)\frac{\partial}{\partial\tau} - \frac{\sinh(\eta)}{\tau}\frac{\partial}{\partial\eta} \\
-\sinh(\eta)\frac{\partial}{\partial\tau} + \frac{\cosh(\eta)}{\tau}\frac{\partial}{\partial\eta} \\
-\partial_R^1\\
-\partial_R^2
\end{array}
\right)\,.\label{gradient}
\eea

The generators $t_a$ of SU$(N_c)$ satisfy
\bea
\label{gen-def}
&& [t_a,t_b] = i f_{abc} t_c  \nonumber\\
&& \text{Tr}(t_a t_b) = \frac{1}{2}\delta_{ab} \nonumber \\
&& f_{abc} = -2i \text{Tr}\big(t_a[t_b,t_c]\big) \,.
\eea
Functions like $A_\mu$, $J_\mu$, $\rho$ and $\Lambda$ are SU$(N_c)$ 
valued functions and can be written as linear combinations of the SU$(N_c)$ generators.
In the adjoint representation we write the generators with a tilde as $(\tilde t_a)_{bc} = -i f_{abc}$. 

The covariant derivative is defined as $D_\mu = \partial_\mu - i  g A_\mu$. In the adjoint representation this becomes $D_{\mu\,ab} = \delta_{ab}\partial_\mu - g f_{abc}A_{\mu\,c}$. Gauge transformations are written 
\bea
\label{gt-def}
&& U(x) = \text{exp}[i t_a \theta_a(x)]\nonumber\\
&& \Psi(x) \to   U^\dagger(x)\Psi(x)  \nonumber\\
&& A^\mu(x) \to \frac{i}{g} U^\dagger(x) \partial^\mu U(x) + U^\dagger(x) A^\mu(x) U(x)  \,\label{gtA}\\
&& D_\mu(x) \to  U^\dagger(x) D_\mu(x) U(x) \label{gtD}\\
&& F_{\mu\nu}(x) \to  U^\dagger(x) F_{\mu\nu}(x)U(x) \,. \label{gtF}
\eea
We will use two specific gauge transformations (see equations (\ref{myU1}, \ref{myU2}))
 \bea
 U_1(x^-,\vec x_\perp) 
 = {\cal P}{\rm exp}\big[i g \int_{-\infty}^{x^-} dz^- \Lambda_{1\,a}(z^-,\vec x_\perp) \,t_a\big]\,\nonumber\\
 U_2(x^+,\vec x_\perp) 
 = {\cal P}{\rm exp}\big[i g \int_{-\infty}^{x^+} dz^+ \Lambda_{2\,a}(z^+,\vec x_\perp)\,t_a\big]\, \label{myU-copy}
 \eea
where we use the ``left later'' convention for path ordering. 
In the adjoint representation we write
 \bea
 W_1(x^-,\vec x_\perp) 
 = {\cal P}{\rm exp}\big[i g \int_{-\infty}^{x^-} dz^- \Lambda_{1\,a}(z^-,\vec x_\perp) \,\tilde t_a\big]\,\nonumber\\
 W_2(x^+,\vec x_\perp) 
 = {\cal P}{\rm exp}\big[i g \int_{-\infty}^{x^+} dz^+ \Lambda_{2\,a}(z^+,\vec x_\perp)\,\tilde t_a\big]\,. \label{myW}
 \eea
 These matrices satisfy the usual identity
 \bea
 U^\dagger t_a U = W_{ab}t_b = t_b W^\dagger_{ba}\,.
 \label{identity}
 \eea
 
Covariant derivatives in Milne coordinates include both the gauge field contribution and 
 curvature terms. 
Products of  covariant derivatives acting on a scalar function $\phi$ are written 
\bea
&& \nabla_\mu \, \phi =D_\mu \, \phi \nn\\
&& \nabla_\mu \nabla_\nu \,\phi = (D_\mu D_\nu  -\Gamma^\lambda_{\mu\nu}D_\lambda) \, \phi \nonumber \\
&& \nabla_\alpha \nabla_\mu \nabla_\nu \,\phi = (D_\alpha \nabla_\mu \nabla_\nu  -\Gamma^\tau_{\alpha\mu} \nabla_\tau \nabla_\nu-\Gamma^\tau_{\alpha\nu}\nabla_\mu \nabla_\tau)\,\phi \,.\label{nabla}
\eea
The connection $\Gamma^\lambda_{\mu\nu}$ can be calculated from the metric tensor (\ref{metric}) and one easily shows that the only non-zero components are
\bea
\label{connect}
\Gamma^0_{11} = \tau \text{~~ and~~} \Gamma^1_{01} = \Gamma^1_{10} = \frac{1}{\tau}\,.
\eea

\section{Energy-momentum tensor in terms of fields}
\label{field-exp-BETA}

Our result for the energy-momentum tensor in equation (\ref{TTmink}) can be written in terms of electric and magnetic fields. 
To obtain this expression we transform the field-strength tensor to Minkowski space and then extract the components of the electric and magnetic fields. 
We remind the reader that in our notation a Minkowski space 4-vector is written $v^\mu_{\rm mink} = (v^0,v^z,v^x,v^y)= (v^0,v^z,\vec v_\perp)$ and the field-strength tensor therefore takes the form 
\bea
\label{F-EB}
F^{\mu\nu}_{\rm mink} = \left[ {\begin{array}{cccc}
   0 & -E^z & -E^x & -E^y \\
   E^z & 0 & -B^y & B^x \\
   E^x & B^y & 0 & -B^z \\
   E^y & -B^x & B^z & 0 \\
  \end{array} } \right]\,.
\eea 
The energy-momentum tensor can then be written in terms of field components using equations (\ref{Tmunu}, \ref{F-EB}). 
Since the energy-momentum tensor is symmetric we only need to give the components on the upper half triangle which are
\bea
&&T_{\rm mink}^{00}=\frac{1}{2}\left( E^x_a E^x_a+ E^y_a E^y_a+E^z_a E^z_a+ B^x_a B^x_a+ B^y_a B^y_a+B^z_a B^z_a\right)\nonumber\\
&& T_{\rm mink}^{01} = E^y_a B^z_a-E^z_a B^y_a\nonumber\\
&&T_{\rm mink}^{02}   =E^z_a B^x_a-E^x_a B^z_a\nonumber\\
&&T_{\rm mink}^{03} = E^x_a B^y_a-E^y_a B^x_a \nonumber \\
&&T_{\rm mink}^{11} = -\frac{1}{2}\left( E^x_a E^x_a-E^y_a E^y_a-E^z_a E^z_a+B^x_a B^x_a-B^y_a B^y_a-B^z_a B^z_a \right)\nonumber \\
&&T_{\rm mink}^{12} = -E^x_a E^y_a-B^x_a B^y_a \nonumber \\
&&T_{\rm mink}^{13} = -E^x_a E^z_a-B^x_a B^z_a \nonumber \\
&&T_{\rm mink}^{22} = \frac{1}{2}(E^x_a E^x_a-E^y_a E^y_a+E^z_a E^z_a+B^x_a B^x_a-B^y_a B^y_a+B^z_a B^z_a) \nonumber\\
&&T_{\rm mink}^{23} = -E^y_a E^z_a-B^y_a B^z_a \nonumber \\
&&T_{\rm mink}^{33} = \frac{1}{2}(E^x_a E^x_a+E^y_a E^y_a-E^z_a E^z_a+B^x_a B^x_a+B^y_a B^y_a-B^z_a B^z_a)\,.
\eea
We then transform our result for the field-strength tensor in Milne coordinates into Minkowski coordinates, and 
then extract the field components. 
All field components at $\tau>0$ can be written in terms of the lowest order components in equation (\ref{Bz-initial}).
We remind the reader that our notation is 
$E \equiv E^z(0,\vec x_\perp)$, $B\equiv B^z(0,\vec x_\perp)$ and ${\cal D}^i = \partial^i-ig \alpha_\perp^i(0,\vec x_\perp)$.
The transverse field components $\vec E_\perp$ and $\vec B_\perp$ have contributions only at orders that correspond to odd powers of $\tau$
and the longitudinal components $E^z$ and $B^z$ get contributions only from even powers of $\tau$. 
Our results are
\bea
&& E_{(1)}^i(\vec x_\perp) = -\frac{1}{2} \sinh(\eta)[{\cal D}^i,E] -\frac{1}{2}\epsilon^{ij}\cosh (\eta)[{\cal D}^j,B] \nn\\
&& B_{(1)}^i(\vec x_\perp) = \frac{1}{2}\epsilon^{ij}\cosh (\eta)[{\cal D}^j,E] -\frac{1}{2} \sinh(\eta)[{\cal D}^i,B] \,. 
\label{fields-order1} \nn\\
&& E_{(2)}^z = \frac{1}{4}[{\cal D}^i,[{\cal D}^i,E]] \nn \\
&& B_{(2)}^z = \frac{1}{4}[{\cal D}^i,[{\cal D}^i,B]] \nn\\[2mm]
&& E^i_{(3)} = -\frac{1}{16}\sinh(\eta)[{\cal D}^i,[{\cal D}^j,[{\cal D}^j,E]]]
- \frac{1}{16}\cosh(\eta)\epsilon^{ij}[{\cal D}^j,[{\cal D}^k,[{\cal D}^k,B]]] \nn\\
&& ~~~~~ 
-\frac{ig}{16}\cosh(\eta)\Big( [E,[{\cal D}^i,E]] - [B,[{\cal D}^i,B]] \Big)
-\frac{ig}{8}\sinh(\eta)\epsilon^{ij}[E,[{\cal D}^j,B]]\nn\\[2mm]
&& B^i_{(3)} = \frac{1}{16}\cosh(\eta)\epsilon^{ij}[{\cal D}^j,[{\cal D}^k,[{\cal D}^k,E]]]
-\frac{1}{16} \sinh(\eta)[{\cal D}^i,[{\cal D}^j,[{\cal D}^j,B]]] \nn\\
&& ~~~~~ 
+\frac{ig}{16}\sinh(\eta)\epsilon^{ij} \Big( [E,[{\cal D}^j,E]]  - [B,[{\cal D}^j,B]] \Big)
-\frac{ig}{8}\cosh(\eta)[E,[{\cal D}^i,B]]\nn\\[2mm]
&& E^z_{(4)} = \frac{1}{64}[{\cal D}^i,[{\cal D}^i,[{\cal D}^j,[{\cal D}^j,E]]]]
 + \frac{ig}{16} \epsilon^{ij}[[{\cal D}^i,E],[{\cal D}^j,B]]  \nn\\[2mm]
&& B^z_{(4)} = \frac{1}{64}[{\cal D}^i,[{\cal D}^i,[{\cal D}^j,[{\cal D}^j,B]]]]
 + \frac{3ig}{64} \epsilon^{ij}[[{\cal D}^i,B],[{\cal D}^j,B]] \nn\\
&& ~~~~~~  - \frac{ig}{64} \epsilon^{ij}[[{\cal D}^i,E],[{\cal D}^j,E]] 
 - \frac{g^2}{64} [E,[E,B]]\,.\label{comps} 
\eea
We comment that these results have somewhat different form compared to Ref. \cite{Chen:2015wia} but are equivalent. 

\section{Initial conditions}
\label{initial-sec}

\subsection{Preliminaries}

In this appendix we will derive the initial conditions for the differential equations that give the gauge potentials in the forward light-cone. 
These conditions were originally obtained by working with sources with zero width across the light-cone that are represented with delta functions, by matching singular terms \cite{Kovner:1995ts, Kovner:1995ja}. We work with sources with a small but finite width, and therefore the boundary conditions should be obtained by integrating the YM equation across the light-cone. 
We start from the YM equation in the adjoint representation  which has the form
\bea
&& \partial_\mu\partial^\mu A^\nu_a-\partial_\mu\partial^\nu A^\mu_a  
 +g f_{abc}\left(
(\partial_\mu A^\mu_b)A^\nu_c - 2(\partial_\mu A^\nu_b)A^\mu_c + (\partial^\nu A^\mu_b)A_{\mu c}
\right)~~~~~~~~~~~~~~~~~  \nn\\ 
&& \hspace{7cm}  +g^2 f_{abc}f_{cmn}A_{\mu b} A^\mu_m A^\nu_n - J^\nu_a = 0 \,.\label{YM-adjoint}
\eea
We can find boundary conditions that relate the ansatz functions in different regions of spacetime by integrating the YM equation across the lines that separate the different regions. 

\subsection{First boundary condition}

In the case of singular sources, the first boundary condition is obtained by matching singular terms in the YM equation at the point $x^+=x^-=0$ at the tip of the light-cone. To obtain the corresponding condition for sources of finite width, we consider the integral of the YM equation over a small diamond shaped area centered on the tip of the light-cone. 
Taking 
$\nu=i$ (one of the transverse spatial indices) we calculate the integral 
\bea
\lim_{{\rm w}\to 0}\int_{-\ww/2}^{\ww/2} dx^-\, \int_{-\ww/2}^{\ww/2} dx^+\,(\text{YM equation}) = 0
\label{ym-in1}
\eea
where the zero on the right of the equation is from the fact that the potentials in all regions of spacetime satisfy the YM equation. 
The contribution to the left side of (\ref{ym-in1}) from most terms in the YM equation is trivially zero, but there are some terms that do not automatically give zero. The conditions that force the sum of these terms to be zero are the boundary conditions we are looking for. 

The densities $\rho_1(x^-,\vec x_\perp)$ and $\rho_2(x^+,\vec x_\perp)$ diverge as $1/{\rm w}$ when 
${\rm w}\to 0$, but the pre-collision potentials in light-cone gauge remain finite, which can be seen from equations (\ref{myU1}-\ref{pg2}).
It is straightforward to show that there is only one term in the integrand that gives a non-zero contribution to the integral. 
This term is $\partial^+\partial^- A^i_a(x^+,x^-,\vec x_\perp)$ and gives
\bea
0 &=&  \int_{-\ww/2}^{\ww/2} dx^-\int_{-\ww/2}^{\ww/2} dx^+ \, \partial^+\partial^- A^i_a(x^+,x^-,\vec x_\perp) \nn\\
&=& A^i_a(\ww/2,\ww/2,\vec x_\perp) - A^i_a(\ww/2,-\ww/2,\vec x_\perp) - A^i_a(-\ww/2,\ww/2,\vec x_\perp) + A^i_a(-\ww/2,-\ww/2,\vec x_\perp)\,. \nn
 \eea
Taking the limit $\ww\to 0^+$ and using equation (\ref{ansatz}) we obtain 
\bea
\alpha_{\perp\,a}^i(0,\vec x_\perp) = \lim_{{\rm w}\to 0}\left(\beta_{1\,a}^i(x^-,\vec x_\perp) + \beta_{2\,a}^i(x^+,\vec x_\perp)\right) \,
\eea
which is the 
first boundary condition in equation (\ref{b-conds}), written in the adjoint representation. 

\subsection{Second boundary condition}

When singular sources are used, the second boundary condition is obtained by matching singular terms in the YM equation across the positive branch of one of the light-cone variable axes. 
To obtain the corresponding condition for sources of finite width, we consider the integral of the YM equation across a small strip centered on the positive $x^-$ axis. 
We set the free index $\nu$ in the YM equation (\ref{YM-adjoint}) to $\nu=-$ and calculate
\bea
\lim_{{\rm w}\to 0}\int_{-\ww/2}^{\ww/2} dx^+\,(\text{YM equation}) = 0\,.
\label{ym-in2}
\eea
The only non-zero contributions come from terms with a derivative with respect to $x^+$. 
Two of these terms that give zero are
\bea
t_1 = \int_{-\ww/2}^{\ww/2} dx^+\, A^+_a(x^+,x^-,\vec x_\perp) \partial^- A_b^-(x^+,x^-,\vec x_\perp) 
\le  \bar A_a^+ \int_{-\ww/2}^{\ww/2} dx^+\, \partial^- A_b^-(x^+,x^-,\vec x_\perp)\nonumber
\eea
\bea
t_2 && = \int_{-\ww/2}^{\ww/2} dx^+\, A_a^-(x^+,x^-,\vec x_\perp) \partial^- A_b^+(x^+,x^-,\vec x_\perp) 
\le  \bar A_a^- \int_{-\ww/2}^{\ww/2} dx^+\, \partial^- A_b^+(x^+,x^-,\vec x_\perp)\, \nn
\eea
where $\bar A_a^+$ and $\bar A_a^-$ indicate the maximum value of the corresponding potential on $x^+\in[-\ww/2,\ww/2]$.
For the first term, the integral is finite but the prefactor goes to zero, using (\ref{ansatz}). For the second term, the the prefactor is finite but the integral goes to zero, again using (\ref{ansatz}). 
Collecting the non-zero terms equation (\ref{ym-in2}) becomes
\bea
0=\int_{-\ww/2}^{\ww/2} dx^+\,&&\bigg[2\,\partial^+\partial^- A_a^-(x^+,x^-,\vec x_\perp)
+\partial^-\partial^i A_a^i(x^+,x^-,\vec x_\perp)  \nn\\
&&
  + g f_{abc} A^i_b(x^+,x^-,\vec x_\perp) \partial^- A^i_c(x^+,x^-,\vec x_\perp) - J^-_a(x^+,\vec x_\perp)\bigg]\,.\label{ic2-a}
\eea
The first term is straightforward to integrate and gives
\bea
\lim_{\ww\to 0}\text{term}_1 = 2\lim_{\ww\to 0}\left(\partial^+ A^-_a(\ww/2,x^-,\vec x_\perp) - \partial^+ A^-_a(-\ww/2,x^-,\vec x_\perp)\right) %
 = -2\alpha(0,\vec x_\perp)\,\label{surv1}
\eea
where we used (\ref{ansatz}) in the last step.
The results of the previous section tell us that 
\bea
A^i_a(x^+,x^-,\vec x_\perp) = \beta_{1\,a}^i(x^-,\vec x_\perp) + \beta_{2\,a}^i(x^+,\vec x_\perp) + {\cal O}(\ww) \text{~~for~~} x^+\in[-\ww/2,\ww/2]\, \label{combo}
\eea
from which we find that the second term of (\ref{ic2-a}) gives
\bea
\lim_{\ww\to 0} \text{term}_2 = \lim_{\ww\to 0} \int_{-\ww/2}^{\ww/2} dx^+ \partial^- \partial^i \beta^i_{2\,a}(x^+,\vec x_\perp)\,.
\eea
The limit of the third term in (\ref{ic2-a}) can be written using equation (\ref{combo}) as
\bea
 \lim_{\ww\to 0}\text{term}_3 
 =  gf_{abc}\lim_{\ww\to 0}\bigg(\beta^i_{1\,b}(x^-,\vec x_\perp) \int_{-\ww/2}^{\ww/2} && dx^+\, \partial^- A^i_c(x^+,x^-,\vec x_\perp) \label{third-term}\\
 + \int_{-\ww/2}^{\ww/2} && dx^+\,\beta^i_{2\,b}(x^+,\vec x_\perp) \partial^- A^i_c(x^+,x^-,\vec x_\perp) \bigg)\,. \nn
\eea
The first term on the right side of (\ref{third-term}) is
\bea
\lim_{\ww\to 0}\text{term}_{3a} &=&   gf_{abc}\lim_{\ww\to 0}\beta^i_{1\,b}(x^-,\vec x_\perp)\left(A^i_c({\rm w}/2,x^-,\vec x_\perp)-\beta^i_{1\,c}(x^-,\vec x_\perp)\right) \,\nn\\
&=& g f_{abc} \,\lim_{{\rm w}\to 0} \beta^i_{1\,b}(x^-,\vec x_\perp)\beta^i_{2\,c}(x^+,\vec x_\perp)\, \label{surv3}
\eea 
where we have used equation (\ref{ansatz}) in the first step and equation (\ref{combo}) to obtain the second line.
The second term on the right side of (\ref{third-term}), using (\ref{combo}), is
\bea
\lim_{\ww\to 0}\text{term}_{3b} &=&  gf_{abc}\lim_{\ww\to 0} \int_{-\ww/2}^{\ww/2} dx^+\, \beta^i_{2\,b}(x^+,\vec x_\perp) \partial^- \beta^i_{2\,c}(x^+,\vec x_\perp)\,.
\eea

Collecting these results equation (\ref{ym-in2}) with $\nu=-$ now has the form
\bea
\lim_{\ww\to 0}\left[\text{term}_1 + \text{term}_2 + \text{term}_{3a} + \text{term}_{3b} - \int_{-\ww/2}^{\ww/2} dx^+ J^-(x^+,\vec x_\perp)\right] = 0\,.
\eea
It is straightforward to show that the piece $\lim_{\ww\to 0}$[$\text{term}_2  + \text{term}_{3b} - \int_{-\ww/2}^{\ww/2} dx^+ J^-(x^+,\vec x_\perp)$] is just the integral of the YM equation for the pre-collision potential $\beta_2$ in the absence of the source corresponding to ion 1, and can therefore be set to zero. 
Using equations (\ref{surv1}, \ref{surv3}) the surviving terms give
\bea
\alpha_a(0,\vec x_\perp) = \frac{g}{2} f_{abc} \,\lim_{{\rm w}\to 0^+} \beta^i_{1\,b}(x^-,\vec x_\perp)\beta^i_{2\,c}(x^+,\vec x_\perp)\,
\eea
which is the second boundary condition in equation (\ref{b-conds}), written in the adjoint representation.

\section{2-potential correlation function}
\label{BBxx}
In this appendix we give the derivation of the 2-point correlation function defined in equation (\ref{core5-20}), the result for which is given in  equations (\ref{B-res}, \ref{Gamma-tilde-def}, \ref{gamma-tilde-def}). 
The result has appeared previously in the literature, and we present it here for completeness, and to explain our notation. 

The pre-collision  potentials can be expressed in terms of the ion sources by solving the YM equation in the pre-collision region.
 This is done most easily by making a gauge transformation. 
The ansatz (\ref{ansatz}) together with the boundary condition (\ref{b-conds}) expresses the pre-collision potentials in terms of the transverse components $\vec\beta^i_1$ and $\vec\beta^i_2$ which are conventionally called light-cone gauge potentials. 
We can transform these potentials without violating our chosen gauge condition (\ref{foch}) by exploiting residual gauge freedom. 
Furthermore, since the two pre-collision regions to the left and right of the forward light-cone are not causally connected, we can work in different gauges in each of these regions. 

First we discuss ion 1. 
The pre-collision potential can be transformed so that the light-cone gauge form
\bea
&& \beta_1^-(x^-,\vec x_\perp) = \beta_1^+(x^-,\vec x_\perp)=0\text{~and~} \beta_1^i(x^-,\vec x_\perp)\ne 0\label{resid-1}
\eea
becomes
\bea
\beta_{1\,{\rm cov}}^-(x^-,\vec x_\perp) = \beta_{1\,{\rm cov}}^i(x^-,\vec x_\perp)=0\text{~and~} \beta_{1\,{\rm cov}}^+(x^-,\vec x_\perp)\equiv \Lambda_1(x^-,\vec x_\perp)\,.\label{resid-1b}
\eea
The new potential satisfies $\partial_\mu\beta_{1\,{\rm cov}}^\mu = \partial^-\beta^+_{1\,{\rm cov}}=0$ and is conventionally called the covariant gauge potential. In covariant gauge the YM equation has the simple form
\bea
&& \nabla_\perp^2 \Lambda_1(x^-,\vec x_\perp) = -\rho_1(x^-,\vec x_\perp) 
\label{YM-cov}
\eea
which can be easily solved to obtain
\bea
\Lambda_1(x^-,\vec x_\perp) = \int d^2z_\perp\,G(\vec x_\perp - \vec z_\perp)\,\rho_1(x^-,\vec z_\perp) 
\label{LAM-def}
\eea
with 
\bea
G(\vec x_\perp) = \frac{1}{2\pi} K_0(m |\vec x_\perp|)\,.
\label{Gf-def}
\eea
The function $K_0$ is a modified Bessel function of the second kind, and $m$ is an infra-red regulator whose definition will be discussed in section \ref{sec-ir-reg}. 
In exactly the same way we obtain the covariant gauge solution for the second ion
\bea
\Lambda_2(x^+,\vec x_\perp) = \int d^2z_\perp\,G(\vec x_\perp - \vec z_\perp)\,\rho_2(x^+,\vec z_\perp)\,.
\label{LAM-def-2}
\eea

Next we must find the residual gauge transformation that allows us to obtain the light-cone gauge potentials from our covariant gauge solutions. For ion 1 we must solve $\beta_1^+(x^-,\vec x_\perp)=0$ with
\bea
\beta_1^+(x^-,\vec x_\perp) = \frac{i}{g} U_1^\dagger(x^-,\vec x_\perp) \partial^+ U_1(x^-,\vec x_\perp) + U_1^\dagger(x^-,\vec x_\perp) \beta_{1\,{\rm cov}}^+(x^-,\vec x_\perp) U_1(x^-,\vec x_\perp) \,.
\eea
The solution is
 \bea
 U_1(x^-,\vec x_\perp) 
 = {\cal P}{\rm exp}\big[i g \int_{-\infty}^{x^-} dz^- \Lambda_1(z^-,\vec x_\perp)\big]\,\label{myU1}
 \eea
where the lower limit on the integral is chosen to give retarded boundary conditions. 
The transverse components in light-cone gauge therefore satisfy 
\bea
\beta_1^i(x^-,\vec x_\perp) = \frac{i}{g} U_1^\dagger(x^-,\vec x_\perp) \partial^i U_1(x^-,\vec x_\perp) \,.
\label{pg1}
\eea
For ion 2 we proceed in the same way. The covariant gauge potential is defined as 
\bea
\beta_{2\,{\rm cov}}^-(x^+,\vec x_\perp) = \beta_{2\,{\rm cov}}^i(x^+,\vec x_\perp)=0\text{~and~} \beta_{2\,{\rm cov}}^+(x^+,\vec x_\perp)\equiv \Lambda_2(x^+,\vec x_\perp)\,,\nn
\eea
the corresponding residual gauge transformation is 
 \bea
 U_2(x^+,\vec x_\perp) 
 = {\cal P}{\rm exp}\big[i g \int_{-\infty}^{x^+} dz^+ \Lambda_2(z^+,\vec x_\perp)\big]\,, \label{myU2}
 \eea
 and the light-cone gauge transverse potential is obtained from the covariant potential as
 \bea
 \beta_2^i(x^+,\vec x_\perp)  =  \frac{i}{g} U_2^\dagger(x^+,\vec x_\perp) \partial^i U_2(x^+,\vec x_\perp)\,. \label{beta-foch}
 \label{pg2}
 \eea
 
A more convenient expression for the light-cone gauge potential can be constructed from these results. For ion 1 we use equations (\ref{resid-1}, \ref{resid-1b}) to obtain
\bea
&& F_1^{+i}(x^-,\vec x_\perp)=\partial^+\beta_1^i(x^-,\vec z_\perp) \nn\\
&& F^{+i}_{1\rm cov}(x^-,\vec x_\perp) = -  \partial_x^i \Lambda_1(x^-,\vec x_\perp) 
\eea
which gives
\bea
\beta_1^i(x^-,\vec x_\perp) &=& \int_{-\infty}^{x^-} dz^- \; F_1^{+i}(z^-,\vec x_\perp) = \int_{-\infty}^{x^-} dz^-  U_1^\dagger(z^-,\vec x_\perp) F_{1\rm cov}^{+i}(z^-,\vec x_\perp) U_1(z^-,\vec x_\perp) \nn\\
&=& -\int_{-\infty}^{x^-} dz^- \; U_1^\dagger(z^-,\vec x_\perp) \partial^i_x\Lambda_1(z^-,\vec x_\perp) U_1(z^-,\vec x_\perp)\nn\\
&=& -\int_{-\infty}^{x^-} dz^-  \; \partial^i_x\Lambda_{1\,a}(z^-,\vec x_\perp) (W_1)_{ab}(z^-,\vec x_\perp) t_b\,,
\label{lc2cov}
\eea
where we use $W$ to denote the Wilson line in the adjoint representation (see Appendix \ref{apA}). An analogous result for the light-cone gauge potential from the second ion can be obtained in the same way and we do not write it explicitly.

%Now we consider correlation functions of pre-collision potentials. 
%As explained in section \ref{sec-corr}, we use the Glasma Graph approximation and therefore we only need to calculate 2-point correlators. 

Now we calculate the correlator in the first line of equation (\ref{core5-20}), and we suppress the index 1 that indicates that potentials and sources are those of the first ion. 
The calculation of correlators for the second ion is exactly analogous. 
Using equation (\ref{lc2cov}) we have that the correlator we want to calculate can be written
\bea
&& \langle \beta^i(x^-,\vec x_\perp) \beta^j(y^-,\vec y_\perp)\rangle \label{core3b} \\
 ~~~ && =  \int_{-\infty}^{x^-} dz^- \int_{-\infty}^{y^-} dw^- \; t_ct_d \langle W_{ec}(z^-,\vec x_\perp) W_{fd}(w^-,\vec y_\perp) 
 \rangle \partial_x^i\partial^j_y \langle  \Lambda_e(z^-,\vec x_\perp) \Lambda_f(w^-,\vec y_\perp) \rangle \,. \nn
\eea
We define the function $\gamma$ using the equation 
\bea
\delta_{ab}\,g^2\, \delta(x^- - y^-)\gamma(x^-,\vec x_\perp,\vec y_\perp)  \equiv  \langle \Lambda_a(x^-,\vec x_\perp)\Lambda_b(y^-,\vec y_\perp)\rangle 
\label{corr1-cov}
\eea
and $\gamma$  is obtained from equations (\ref{LAM-def}, \ref{big-in}) as
\bea
&& \gamma(x^-,\vec x_\perp,\vec y_\perp) = \int d^2 \vec z_\perp 
\, \lambda(x^-,\vec z_\perp)\, G(\vec x_\perp-\vec z_\perp)\,G(\vec y_\perp- \vec z_\perp)\,.
\label{gamma-def-0}
\eea 
The correlator of Wilson lines has been calculated in Ref. \cite{Fukushima:2007dy} where it is shown that 
 \bea
 \label{WW-res-copy}
&& \delta_{cd}\,\langle  W_{a c}(x^-,\vec x_\perp) W_{b d}(y^-,\vec y_\perp)  \rangle \\
&&  = \delta_{ab}{\rm exp}\left[\frac{g^4 N_c}{2}\int^{x^-}_{-\infty} dz^-\big(2\gamma(z^-,\vec x_\perp,\vec y_\perp) - \gamma(z^-,\vec x_\perp,\vec x_\perp) - \gamma(z^-,\vec y_\perp,\vec y_\perp)\big)\right]\,.\nonumber
 \eea
Using  equations (\ref{corr1-cov}, \ref{WW-res-copy}) we find that equation (\ref{core3b}) can be written in the adjoint representation as
 \bea
 \langle \beta_a^i(x^-,\vec x_\perp) \beta_b^j(y^-,\vec y_\perp)\rangle 
&=& \delta_{ab} \,g^2  \int^{x^-}_{-\infty}dz^-\int^{y^-}_{-\infty}dw^-
\,\delta(z^--w^-) \;\partial_x^i \partial_y^j \gamma(z^-,\vec x_\perp,\vec y_\perp) \nn \\[2mm]
{\rm exp}\bigg[
\frac{g^4 N_c}{2}\int^{z^-}_{-\infty} dv^-  \hspace*{-.5cm} && \big(2\gamma(v^-,\vec x_\perp,\vec y_\perp) - \gamma(v^-,\vec x_\perp,\vec x_\perp) - \gamma(v^-,\vec y_\perp,\vec y_\perp)\big)
\bigg]  \,.
\label{core3-adjoint}
 \eea
Next we take the limit that the width  of the source current $\rho(x^-,\vec x_\perp)$ across the light-cone
goes to zero. 
Using equation (\ref{norm}) we rewrite (\ref{gamma-def-0}) as
\bea
&& \gamma(x^-,\vec x_\perp,\vec y_\perp) = h(x^-) \,\tilde \gamma(\vec x_\perp,\vec y_\perp)\,
\eea
where 
\bea
&& \tilde \gamma(\vec x_\perp,\vec y_\perp) = \int d^2 z_\perp \, \mu(\vec z_\perp)\, G(\vec x_\perp-\vec z_\perp)\,G(\vec y_\perp-\vec z_\perp) \,.
\label{gamma-tilde-def-copy}
\eea
We  define the functions $\Gamma (\vec x_\perp,\vec y_\perp)$ and $\tilde\Gamma (\vec x_\perp,\vec y_\perp)$ as
\bea
&&  \Gamma(z^-,\vec x_\perp,\vec y_\perp)  =  2\gamma(z^-,\vec x_\perp,\vec y_\perp) - \gamma(z^-,\vec x_\perp,\vec x_\perp) - \gamma(z^-,\vec y_\perp,\vec y_\perp) \nonumber\\
&& \tilde\Gamma (\vec x_\perp,\vec y_\perp) = 2\tilde\gamma(\vec x_\perp,\vec y_\perp) - \tilde\gamma(\vec x_\perp,\vec x_\perp) - \tilde\gamma(\vec y_\perp,\vec y_\perp)  \,. \label{Gamma-tilde-def-copy}
\eea
Using these definitions we rewrite the exponential in (\ref{core3-adjoint}) as 
\bea
  {\rm exp}\left[~ \cdots ~\right] =  {\rm exp}\left[\frac{g^4 N_c}{2}\;\tilde\Gamma(\vec x_\perp,\vec y_\perp) \int^{z^-}_{-\infty} dv^- h(v^-)  \right] \,. \label{core4}
 \eea
We need to substitute (\ref{core4}) into (\ref{core3-adjoint}) and take the limit that the width of the source distributions goes to zero. The function $h(z^-)$ behaves like a delta function in this limit, and it appears therefore that the calculation should be simple. However, we must proceed carefully to be sure that the delta functions in the integrand have support. 
We define
 \bea
f(x^-) \equiv  \int^{x^-}_{-\infty} dz^- h(z^-) \label{fg-def} \text{~~and~~} g(\vec x_\perp,\vec y_\perp) \equiv \frac{g^4 N_c}{2}\;\tilde\Gamma(\vec x_\perp,\vec y_\perp)    
\eea
so that equation (\ref{core3-adjoint}) becomes
\bea
&&\langle \beta_a^i(x^-,\vec x_\perp) \beta_b^j(y^-,\vec y_\perp)\rangle = \delta_{ab}\,g^2\,  I(x^-,y^-,\vec x_\perp,\vec y_\perp) \,
 \partial_x^i \partial_y^j \tilde\gamma(\vec x_\perp,\vec y_\perp) \label{core4b}
\eea
where
\bea I(x^-,y^-,\vec x_\perp,\vec y_\perp) = 
&&\int^{x^-}_{-\infty}dz^-\int^{y^-}_{-\infty}dw^-\,\delta(z^--w^-) h(z^-) e^{g(\vec x_\perp,\vec y_\perp) f(z^-)} \nonumber\\
&& =\frac{1}{g(\vec x_\perp,\vec y_\perp)} \int^{x^-}_{-\infty}dz^-\int^{y^-}_{-\infty}dw^-\,\delta(z^--w^-)\frac{\partial}{\partial z^-} e^{g(\vec x_\perp,\vec y_\perp) f(z^-)}\nonumber \\
&& =\frac{1}{g(\vec x_\perp,\vec y_\perp)} \int^{{\rm min}(x^-,y^-)}_{-\infty}dz^- \frac{\partial}{\partial z^-} e^{g(\vec x_\perp,\vec y_\perp) f(z^-)} \nonumber\\
&& =\frac{1}{g(\vec x_\perp,\vec y_\perp)}\big[ e^{g(\vec x_\perp,\vec y_\perp) f({\rm min}(x^-,y^-))}-1\big] \,.
\eea
Taking the limit that the width of the function $h(z^-)$ goes to zero so that $f\big({\rm min}(x^-,y^-)\big)\to 1$ [see equations (\ref{norm}, \ref{fg-def})]
 gives 
\bea \lim_{\ww\to 0}I(x^-,y^-,\vec x_\perp,\vec y_\perp) = \frac{1}{g(\vec x_\perp,\vec y_\perp)}\big[ e^{g(\vec x_\perp,\vec y_\perp) }-1\big] \,.
\label{Ilim}
\eea

Combining equations (\ref{fg-def}, \ref{core4b},  \ref{Ilim}) we write 
 \bea
\lim_{{\rm w} \to 0} \langle \beta_{a}^i(x^-,\vec x_\perp) \beta_{b}^j(y^-,\vec y_\perp)\rangle \equiv \delta_{ab}  B^{ij}(\vec{x}_\perp,\vec y_\perp) 
\, \label{core5-20-copy}
\eea
with
  \bea
&& B^{ij}(\vec{x}_\perp,\vec y_\perp) = \frac{2 }{g^2 N_c \tilde\Gamma(\vec x_\perp,\vec y_\perp)}  \; \left({\rm exp}[\frac{g^4 N_c}{2}\;\tilde\Gamma(\vec x_\perp,\vec y_\perp) ]-1\right)\;
\partial_x^i \partial_y^j \tilde\gamma(\vec x_\perp,\vec y_\perp) \,.\label{B-res-copy}
\eea
The calculation of correlators for the second ion is exactly analogous and equations (\ref{core5-20-copy}, \ref{B-res-copy}) can be used for either the first or the second ion by using the charge density $\mu_1(\vec x_\perp)$ or $\mu_2(\vec x_\perp)$ in equations (\ref{gamma-tilde-def-copy}, \ref{Gamma-tilde-def-copy}).

\end{document}